\documentclass[notoc]{JHEP3}
\pdfoutput=1
\usepackage[pdftex]{graphicx}

\newcommand{\cs}{\langle\sigma_A\upsilon\rangle}

\title{Implications of the Fermi-LAT diffuse gamma-ray measurements on annihilating or decaying Dark Matter}

\author{Gert H\"utsi\\ Tartu Observatory, T\~oravere 61602, Estonia\\ E-mail: \email{gert@aai.ee}}
\author{Andi Hektor\\ National Institute of Chemical Physics and Biophysics, Tallinn 10143, Estonia\\ E-mail: \email{andi.hektor@cern.ch}}
\author{Martti Raidal\\ National Institute of Chemical Physics and Biophysics, Tallinn 10143, Estonia,  \\
Department of Physics, P.O.Box 64, FIN-00014 University of Helsinki, Finland\\ E-mail: \email{martti.raidal@cern.ch}}

\abstract{{\small 
We analyze the recently published Fermi-LAT diffuse gamma-ray measurements in the context of leptonically annihilating or decaying dark matter (DM) with the aim to explain simultaneously the isotropic diffuse gamma-ray and  the PAMELA, Fermi and HESS (PFH) anomalous $e^\pm$ data. 
Five different DM annihilation/decay channels $2e$, $2\mu$, $2\tau$, $4e$, or $4\mu$ (the latter two via an intermediate light particle $\phi$) are generated with PYTHIA. We calculate both the Galactic and extragalactic prompt and inverse Compton (IC) contributions to the resulting gamma-ray spectra. 
To find the Galactic IC spectra we use the interstellar radiation field model from the latest release of GALPROP. For the extragalactic signal we show that the amplitude of the prompt gamma-emission is very sensitive to the assumed model for the extragalactic background light. For our Galaxy we use the Einasto, NFW and cored isothermal DM density profiles and include the effects of DM substructure assuming a simple subhalo model. Our calculations show that for the annihilating DM the extragalactic gamma-ray signal can dominate only if rather extreme power-law concentration-mass relation $C(M)$ is used, while more realistic $C(M)$ relations make the extragalactic component comparable or subdominant to the Galactic signal. For the decaying DM the Galactic signal always exceeds the extragalactic one. In the case of  annihilating DM the PFH favored parameters can be ruled out by gamma-ray constraints only if power-law $C(M)$ relation is assumed.  For DM decaying into $2\mu$ or $4\mu$ the PFH favored DM parameters are not in conflict with the gamma-ray data.  We find that, due to the (almost) featureless Galactic IC spectrum and the DM halo substructure, annihilating DM may give a good simultaneous fit to the isotropic diffuse gamma-ray and to the PFH $e^\pm$ data without being in clear conflict with the other Fermi-LAT gamma-ray measurements.
}}
\preprint{0000.0000}

\begin{document}
\linespread{1.05}

\section{Introduction}
During the last few years several experiments have shown an anomalous excesses in the cosmic electron and positron spectra. The PAMELA satellite has observed a steep rise of positron fraction $e^+/(e^-+e^+)$ at energies above 10 GeV with no significant excess in the cosmic antiproton flux \cite{Adriani:2008zr,Adriani:2008zq}. The Fermi satellite and the HESS atmospheric Cherenkov telescope have measured an excess of high-energy $(e^-+e^+)$ flux with a cut-off of around 800 GeV \cite{Abdo:2009zk,Aharonian:2009ah}. The ATIC and PPB-BETS balloon measurements indicate a similar excess \cite{Chang:2008zzr,Torii:2008xu}. Most excitingly, the excess might originate from the annihilation or decay of the dark matter (DM) particles. The nature of those signatures requires the properties of DM to deviate strongly from the standard freeze-out predictions. The thermally averaged DM annihilation cross-section $\cs$ has to be boosted some orders of magnitude over the standard freeze-out value  $\cs_{\rm std} \simeq 3 \times 10^{-26}$ cm$^3$s$^{-1}$, which might be achieved, e.g., through the Sommerfeld effect~\cite{Hisano:2003ec} (see, e.g., \cite{Cirelli:2008pk,ArkaniHamed:2008qn,Slatyer:2009vg} for the related phenomenological studies) or through the Breit-Wigner resonant enhancement \cite{Feldman:2008xs,Ibe:2008ye,Guo:2009aj,Bi:2009uj}.  On the other hand, the decaying DM \cite{Buchmuller:2007ui,Ibarra:2008jk,Nardi:2008ix,Arvanitaki:2008hq,Ibarra:2009dr} can explain the excess independent of the freeze-out constraints. In both cases, the annihilation or decay of DM should favorably occur only through the leptonic channels~\cite{Cholis:2008hb,Cirelli:2008pk,Donato:2008jk,Cholis:2008qq,Cholis:2008wq}, as no excess in the hadronic channels has been observed. Alternatively, the excess of $e^+$ can potentially be explained by modifying or adding astrophysical sources, e.g. pulsars \cite{Hooper:2008kg,Yuksel:2008rf,Profumo:2008ms,Malyshev:2009tw}.

The high energy leptons of the DM annihilation/decay are inevitably accompanied by the gamma-rays due to the final state radiation of charged leptons and decays of subproducts (``prompt gamma-rays'') and due to the upscattered background photons from the inverse Compton (IC) scattering (``IC gamma-rays'').
Thus, the observed gamma-ray fluxes strongly constrain the above mentioned cosmic ray anomalies from DM annihilation/decay. The strongest gamma-ray constraints should arise from the observations of the Galactic center (GC) \cite{Bergstrom:1997fj,Dodelson:2007gd}, as the density of DM is very high and it is relatively close to us. Those analyses take into account both the prompt and IC gamma-ray contributions \cite{Bertone:2008xr,Bergstrom:2008ag,Bell:2008vx,Cirelli:2009vg,Meade:2009iu,Cholis:2009gv}. On the other hand, the GC is densely populated by different astrophysical objects, which contaminate the gamma-ray signal and introduce significant uncertainty in the derived constraints.  The Galactic gamma-ray signal of DM annihilation/decay at higher latitudes is considerably weaker than the signal from the GC. Despite being considerably weaker, the suppressed contamination by the Galactic astrophysical sources partially compensates the weakness of the signal. 
In addition, adding the Galactic DM substructure into the picture \cite{Diemand:2007qr,Springel:2008by,Kuhlen:2009is,Kamionkowski:2010mi,Cline:2010ag} may significantly change both the magnitude and the morphology of the induced gamma-ray signal as the diffuse signal from DM subhalos can be essentially isotropic. 
The Galactic gamma-ray signal of DM annihilation/decay at higher latitudes can also include a considerable extragalactic contribution. It can originate from different sources: DM annihilation/decay in cosmological distances, active galactic nuclei (AGN), structure formation shocks, starburst galaxies, etc.

Fermi-LAT collaboration has also derived constraints on the annihilating DM properties from the Galactic dwarf spheroidal galaxies \cite{Abdo:2010ex} 
(see \cite{Essig:2009jx,PalomaresRuiz:2010pn} and references therein for the other recent studies). 
Non-detection of gamma-ray signal towards several nearby galaxy clusters has a potential to severely constrain the annihilating DM models \cite{Pinzke:2009cp,Yuan:2010gn}. However, the present Fermi results \cite{FermiDM:2010aa} still allow for the DM parameters giving the best fit to the PAMELA, Fermi, and HESS (PFH) $e^\pm$ anomalies even if the smallest DM substructures are present. 
In addition to gamma-rays, the energetic leptons from DM decay/annihilation produce other accompanying signatures: synchrotron radiation \cite{Bertone:2008xr,Bringmann:2009ca,Crocker:2010gy}, and neutrinos from $\mu$ or $\tau$ decay \cite{Desai:2004pq,Covi:2008jy,Hisano:2008ah,Liu:2008ci,Mandal:2009yk}. The gamma-rays ionize and heat the intergalactic gas. The additional electrons released in the ionization process change the scattering optical depth of the CMB photons \cite{Padmanabhan:2005es,Mapelli:2006ej,Zhang:2006fr,Belikov:2009qx,Galli:2009zc,Slatyer:2009yq,Huetsi:2009ex,Cirelli:2009bb,Kanzaki:2009hf,Yuan:2009xq,Iocco:2009ch,Natarajan:2010dc}. 
Also, the energetic $e^\pm$ released in the annihilation/decay process induce the nonthermal component to the Sunyaev-Zel'dovich effect in galaxy clusters \cite{Colafrancesco:2005ji,2009JCAP...10..013Y,2010JCAP...02..005L}. The energy injection from the DM sector modifies the accurately calculable\footnote{See, e.g., \cite{RubinoMartin:2009ry,Chluba:2010fy} for the latest developments.} standard cosmic recombination process (e.g. \cite{Chen:2003gz,Padmanabhan:2005es,Chluba:2010aa}) and the Big Bang Nucleosynthesis (e.g. \cite{Hisano:2008ti,Hisano:2009rc,Jedamzik:2009uy}).

As in, e.g. Regis \& Ullio \cite{Regis:2009md}, in this study we focus on the gamma-ray signal of DM annihilation/decay at higher Galactic latitudes. Fermi-LAT collaboration published recently their measurement of the diffuse gamma-ray emission at several Galactic regions and their estimation of the ``extragalactic diffuse gamma-ray background'' \cite{collaboration:2010nz}. The published spectra are more constraining and have considerably smaller error bars than the older analogous spectra from the EGRET experiment \cite{Sreekumar:1997un,Strong:2004ry}. The main aim of this paper is to find  out whether the new Fermi-LAT diffuse gamma-ray
measurements can rule out or, instead, to support the PAMELA, Fermi, and HESS  favored models of DM annihilation/decay as given in \cite{Meade:2009iu}\footnote{For an earlier analysis, similar to \cite{Meade:2009iu}, see \cite{Bergstrom:2009fa}.}. 

Our study is ``model independent'': we assume that DM particles annihilate/decay into Standard Model charged leptons, $\ell^{-},\ell^{+}$;  $\ell=e,\,\mu,\,\tau,$ extending the analyses presented in series of works, e.g. \cite{Cirelli:2008pk,ArkaniHamed:2008qn,Cirelli:2009vg,Meade:2009iu,Borriello:2009fa,Papucci:2009gd,Cirelli:2009dv,Strumia:2010zz}. We analyze five channels presented in Table~\ref{tab:channels}. Our general formalism for the extragalactic gamma-rays is presented in our previous work \cite{Huetsi:2009ex}. 
The new ingredient  in this paper is to
include the effect of the absorption of extragalactic gamma-rays due to the pair production on extragalactic background light\footnote{See \cite{Ibarra:2009nw}, where the authors have included the background light while discussing the prospects for detecting the decaying DM with the Fermi-LAT.}.  For that we test different models for the extragalactic background light and show that the 
signal of prompt gamma-rays from the DM annihilation/decay can be significantly reduced in some cases.  For calculation of the Galactic IC spectra we use the interstellar radiation field (ISRF) model from the latest release of GALPROP \cite{Porter:2005qx}. We show that inclusion of the realistic ISRF including
CMB, infrared radiation from dust, and stellar light backgrounds are complicated enough to smear out any spectral feature of the individual components and the resulting spectrum is essentially power-law-like.
To model the Galactic DM halo we use three density profiles: Einasto, NFW and cored isothermal. We first perform our analyses assuming 
that our Galaxy consist of one structureless DM halo with the given density profile. This is probably unrealistic but often used approximation.  After that we repeat our study using a more realistic model including DM subhalos. The technical details of our analyses are collected in Appendices A, B, C.

After Fermi collaboration has made their data publicly available, 
several papers have appeared that estimate the constraints on annihilating/decaying DM properties \cite{Papucci:2009gd,Cirelli:2009dv,Chen:2009uq,Zhang:2009ut,Abazajian:2010sq,Abdo:2010dk}. The first two papers \cite{Papucci:2009gd,Cirelli:2009dv} use the preliminary Fermi-LAT data. In both of these studies the authors have neglected the potential extragalactic contribution while deriving the constraints for the annihilating DM. In \cite{Cirelli:2009dv} the approximate ISRF of our Galaxy is used. However, we show in this paper that the use of more realistic ISRF can significantly change the (line of sight) integrated IC spectra and in some cases the extragalactic diffuse gamma-ray signal may dominate over the Galactic one.  Abazajian et al. \cite{Abazajian:2010sq} use the results of \cite{collaboration:2010nz} but consider only the prompt gamma-ray spectra to constrain their DM models. This is clearly unsatisfactory because, as we show in agreement with \cite{Abdo:2010dk}, in general the bounds are dictated by the IC contribution to the diffuse gamma-ray spectrum while the prompt contribution plays just a subdominant role. The two-peak structure of the diffuse gamma-ray spectrum in DM annihilations into $2\mu$ channel presented in Fig. 4 of  Abdo et al. \cite{Abdo:2010dk} 
also shows that those authors do not use realistic ISRF to calculate the IC spectrum but approximate it with the CMB component.  
As we show in this paper, this approximation does not qualitatively change the bounds derived in that paper but 
may qualitatively affect the whole interpretation of the Fermi-LAT experimental results.
In addition,  both Abazajian et al. \cite{Abazajian:2010sq} and Abdo et al. \cite{Abdo:2010dk} study only  the annihilating DM.
In this work, similarly to  Refs.~\cite{PalomaresRuiz:2010pn,Boehm:2010qt}, we study the difference between annihilating and decaying DM gamma-ray
 signals.

The most important new result of this work is  that the inclusion of realistic IRSF and Galactic DM subhalo structure into the
analyses opens up a new interpretation of the Fermi-LAT isotropic diffuse gamma-ray measurement.  We find that 
$(i)$ the leptonically annihilating DM models can give  good fits to the Fermi-LAT isotropic diffuse gamma-ray data if the extragalactic contribution is subdominant
compared to the Galactic one;
$(ii)$ the best-fit regions of the preferred DM mass and annihilation cross section has an overlap with the best-fit regions of the PFH anomaly  \cite{Meade:2009iu} 
if the boost factor from the Galactic DM substructure is $B_{{\rm sub}}\sim {\cal O}(10)$;
$(iii)$ in the case of decaying DM the fits to the isotropic diffuse data are worse and there is no overlap with the best-fit regions of the PFH anomaly. 
This result implies that a significant fraction of the isotropic diffuse gamma-ray signal observed by Fermi-LAT may actually be of the
Galactic origin. A generic issue that for a good fit the central regions of our Galaxy should give less prominent contribution to the DM annihilation signal than generally expected is (partially) solved with the DM halo substructure.
The simultaneous  fitting of PFH electron/positron and Fermi-LAT diffuse gamma ray data gives a preference to the DM 
models with Sommerfeld enhancement due to a new light intermediate particle $\phi$ \cite{ArkaniHamed:2008qn} (see Table~\ref{tab:channels}). 
If the intermediate particle $\phi$ is long-lived \cite{Rothstein:2009pm}, a scenario not considered in this work, 
their long diffusion length further smears any DM annihilation signal and helps to resolve potential conflicts with the diffuse gamma-ray data measurements from the central Galactic regions.

Our paper is organized as follows. In Section 2 we describe the energy and particle input from the DM annihilation/decay. In Section 3 we solve the radiative transfer equation for the Galaxy and the intergalactic medium. Section 4 presents the constraints as inferred from the Fermi-LAT diffuse gamma-ray data. Section 5 discusses the possibility of fitting the models to the Fermi-LAT isotropic diffuse data. Our summary and final discussion is given in Section 6.

\section{Energy input from DM annihilation or decay}\label{sec1}
In this paper we consider the DM annihilation and decay to Standard Model charged leptons presented in Table~\ref{tab:channels}. In addition to the three two-lepton channels, which were also investigated in our previous work \cite{Huetsi:2009ex}, we have now included two additional four-lepton annihilation and decay channels. In our calculations the intermediate particle $\phi$ has a mass $m_\phi = 1$ GeV, which forbids $\phi$ to decay into $\tau$ leptons, and thus we do not consider a $4\tau$ channel. The decay of $\phi$ (assumed to be prompt) and the final state distributions of $e^\pm$, photons and $\nu$ from the primary two-body states produced in DM annihilations are computed using  PYTHIA 8.1 Monte Carlo tool \cite{Sjostrand:2007gs}. Our DM annihilation scenarios for $2e$, $2\mu$ and $2\tau$ are the same as presented in \cite{Cirelli:2008pk}. The energy partition between three main products of the annihilation/decay process: (i) electrons/positrons, (ii) photons, and (iii) neutrinos/antineutrinos is given in Table~\ref{tab:energy}.
 In channels involving unstable muons and taus a large fraction of  energy is carried away by neutrinos which is lost for the creation processes  of the gamma-ray background. The energetic electrons and positrons instantaneously interact with the ambient photon fields and  create a soft part of the output gamma-ray spectrum via the inverse Compton (IC) mechanism. The most energetic part of the gamma spectrum  consists of the prompt (or direct) photons originating directly from the annihilation/decay process. Table~\ref{tab:energy} shows that the $2\tau$ channel has a particularly high fraction of energetic   prompt photons. Finally, it is important to mention that the spectra from annihilation and decay are related to each other in a simple manner: annihilation of two DM particles with a mass $m_{DM}$ can be seen as a decay of a particle with mass $2m_{DM}$, i.e. all of the decay spectra are indistinguishable from the annihilation spectra for two times smaller particle mass. 

\TABULAR[h]
{c|c|c}
{Annihilation & Decay & Notation \\
\hline
\hline
$2 DM \to e^+ + e^-$ & $DM \to e^+ + e^-$ & $2e$\\
\hline
$2 DM \to \mu^+ + \mu^-$ & $DM \to \mu^+ + \mu^-$ & $2\mu$\\
\hline
$2 DM \to \tau^+ + \tau^-$ & $DM \to \tau^+ + \tau^-$ & $2\tau$\\
\hline
$2 DM \to 2 \phi$, $\phi \to e^+ + e^-$ & $DM \to 2 \phi$, $\phi \to e^+ + e^-$ & $4e$\\
\hline
$2 DM \to 2 \phi$, $\phi \to \mu^+ + \mu^-$ & $DM \to 2 \phi$, $\phi \to \mu^+ + \mu^-$ & $4\mu$}
{The annihilation and decay channels of DM and their notation in the paper.\label{tab:channels}}

\TABULAR[h]
{c|c|c|c}
{channel & $e^\pm$ & photons & neutrinos \\
\hline
\hline
$2 e$ & $\sim 96-97\,\%$ & $\sim 3-4\,\%$ & $0\,\%$\\
\hline
$2 \mu$ & $\sim 34\,\%$ & $\sim 2-3\,\%$ & $\sim 63-64\,\%$\\
\hline
$2 \tau$ & $\sim 16\,\%$ & $\sim 16-17\,\%$ & $\sim 67-68\,\%$\\
\hline
$4 e$ & $\sim 98\,\%$ & $\sim 2\,\%$ & $0\,\%$\\
\hline
$4 \mu$ & $\sim 35\,\%$ & $\sim 1\,\%$ & $\sim 64\,\%$}
{Energy partition between $e^\pm$, photons and neutrinos/antineutrinos for all the studied annihilation and decay channels assuming $m_{DM}=100$ GeV - $10$ TeV.\label{tab:energy}}

\section{Gamma-ray spectra from DM annihilation/decay}
In this section we briefly describe the machinery used to calculate the gamma-ray spectra. We start by introducing the approach used for calculating the contribution from our Galaxy's DM halo, and in the second half of the section we focus on extragalactic part. 
\subsection{Galactic halo contribution}
As described above, the energetic $e^\pm$ created in the annihilation/decay interact with the ambient photon fields through the IC mechanism, where the soft input photons are boosted to gamma-ray energy range. Inside the Galaxy the interstellar radiation field (ISRF) relevant for the IC mechanism has three main components: (i) stellar light, (ii) infrared light, which is the reprocessed form of the stellar radiation by the surrounding dust, (iii) cosmic microwave background (CMB) radiation. In our calculations we take the ISRF from the latest release of GALPROP\footnote{http://galprop.stanford.edu/} \cite{Porter:2005qx}. Since our input $e^\pm$ can be very energetic (we consider $m_{DM}$ up to $10$ TeV) the IC calculations involving stellar and infrared components of the ISRF need to exploit the full energy dependent Klein-Nishina form for the cross-section. For our IC calculations we use the formalism in Blumenthal \& Gould \cite{1970RvMP...42..237B}, and in Cirelli \& Panci \cite{Cirelli:2009vg}\footnote{For the sake of completeness we have given a "quick recipe" for the IC calculations in Appendix~\ref{appa}. Also, Ref. \cite{Colafrancesco:2005ji} presents the IC formalism.}. We have neglected the effect of electron diffusion, i.e. our IC photons are created ``on spot''\footnote{For more rigorous treatment one should first solve the diffusion-loss equation for $e^\pm$ and then calculate the IC spectrum using this new electron distribution.}.  Once the gamma photon is produced, whether directly from the annihilation/decay event or through the IC mechanism, it can propagate almost freely inside the Galaxy, i.e. inside the Galaxy at gamma-ray energy range we can safely neglect any absorption. Thus, to calculate the gamma-ray intensity towards some Galactic latitude $b$ and longitude\ $l$ one just has to sum up the emissivity $j_E$ $[$eV~cm$^{-3}$s$^{-1}$eV$^{-1}]$ along the line of sight
\begin{equation}
I_E(b,l)=\frac{1}{4\pi}\int_{{\rm l.o.s.}}j_E(b,l,r){\rm d}r\,.
\end{equation}
\FIGURE{
\includegraphics[width=1.0\textwidth]{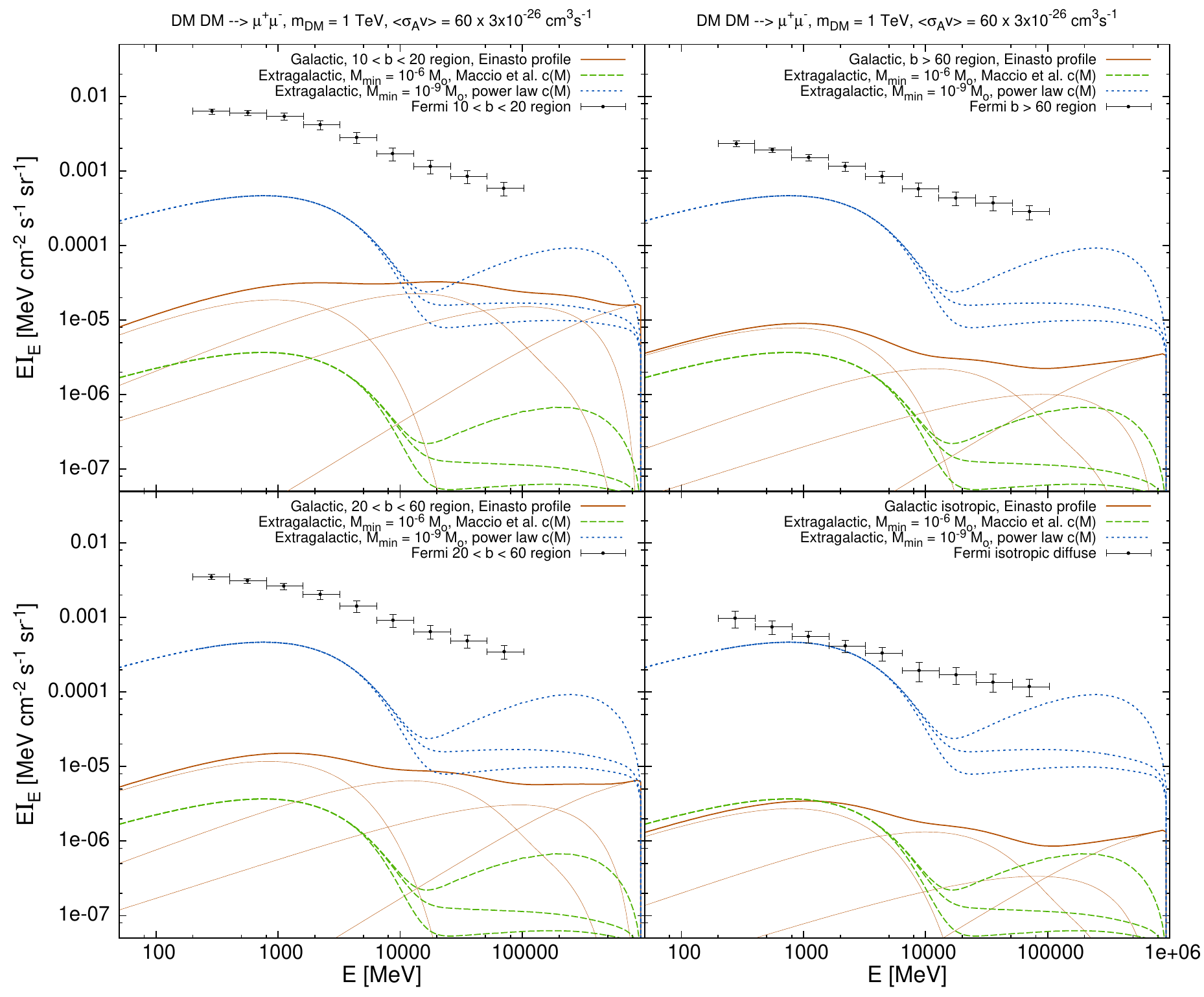}
\caption{Galactic gamma-ray spectra (solid lines) for $2\mu$ annihilation channel with Einasto density profile, $m_{DM}=1$ TeV and with annihilation cross-section $\cs= 60 \times 3 \times 10^{-26}$ cm$^3$s$^{-1}$ for $10<b<20$ (upper left panel), $20<b<60$ (lower left panel), and $b>60$ (upper right panel) regions. Spectrum of the Galactic isotropic component is given on the lower right-hand panel. The decomposition of the Galactic spectra into main components (in increasing order of energy): (i) IC from CMB, (ii) IC from infrared radiation, (iii) IC from stellar light, (iv) prompt emission, is also presented. Dashed and dotted lines represent extragalactic spectra for the Maccio et al. \cite{2008MNRAS.391.1940M} $C(M)$ relation with $M_{\min}=10^{-6}M_{\odot}$ and for the power-law $C(M)$ with $M_{\min}=10^{-9}M_{\odot}$, respectively. The high-energy splitting of these curves corresponds to the three choices for the extragalactic ultraviolet (UV) background (from top to bottom): (i) ``no UV'', (ii) ``realistic UV'', (iii) ``optimistic UV''. The points with errorbars give the Fermi-LAT measurements.}
\label{fig1}}
The ISRF from GALPROP code is given as a grid in Galactic cylindrical coordinates $R$ and $z$, covering the ranges $R=0.25-20.25$ kpc and $z=-5-5$ kpc with a total of $101 \times 41$ grid points. We calculate the emission coefficients $j_E$ for each of the grid points\footnote{In reality we can discard half of the grid points due to the reflection symmetry of the ISRF with respect to the $z=0$ plane.} by looking first at annihilating DM models with a standard thermal production cross section $\cs_{\rm std}\simeq3\times10^{-26}$ cm$^3$s$^{-1}$ and fixing the DM density globally to $\rho_{\odot} = 0.4$ GeV cm$^{-3}$, which is a recently preferred value for the DM density in our local Galactic neighborhood \cite{Catena:2009mf,2010arXiv1003.3101S}. We perform this calculation for all five annihilation channels and for $20$ logarithmically-spaced DM particle masses between $100$ GeV and $10$ TeV. Our emission coefficients consist of four parts: (i) IC from stellar light, (ii) IC from infrared radiation, (iii) IC from CMB, and (iv) prompt emission part. Outside the cylindrical region where the GALPROP ISRF is defined, we continue with two components: (i) IC from CMB, and (ii) prompt emission. This calculation is numerically the most demanding part of the current work. However, it is easy to realize that it can be done only for once and for all: e.g., if one is interested in decaying DM models with a half-life $\tau$, then one can simply modify the emission coefficients $j_E$ by multiplying with a factor of $2m_{DM}/(\tau\cs_{\rm std}\rho_{\odot})$ \cite{Kadastik:2009ts}. Also, it is trivial to rescale emission coefficients in order to accommodate various density profiles for the Galactic halo. As in Cirelli et al. \cite{Cirelli:2009dv} and in Meade et al. \cite{Meade:2009iu}, in our calculations we also use three distinct forms for the Galactic DM density profile \cite{Einasto:1965aa,Navarro:1996gj,Bahcall:1980fb}:
\begin{equation}\label{eq_3.2}
\frac{\rho(r)}{\rho_{\odot}}= \left\{ 
\begin{array}{lll}
\exp\left( -\frac{2}{\alpha}\left[\left(\frac{r}{r_s}\right)^{\alpha}-\left(\frac{r_{\odot}}{r_s}\right)^{\alpha}\right]\right) & {\rm \quad Einasto,}\ r_s=20\ {\rm kpc},\alpha=0.17\,,\\
\frac{r_{\odot}}{r} \left (1 + \frac{r_{\odot}}{r_s}\right)^2 / \left (1 + \frac{r}{r_s}\right)^2 & {\rm \quad NFW,}\ r_s=20\ {\rm kpc}\,,\\
\left(1 + \left [\frac{r_{\odot}}{r_s}\right]^2 \right) / \left(1 + \left [\frac{r}{r_s}\right]^2\right) & {\rm \quad cored\ isothermal,}\ r_s=5\ {\rm kpc}\,,
\end{array} \right.
\end{equation}
where the DM density in our local neighborhood $\rho_{\odot}=\rho(r=r_{\odot}\simeq 8.5\,{\rm kpc})=0.4$ GeV cm$^{-3}$, e.g. \cite{Catena:2009mf,2010arXiv1003.3101S}.

\FIGURE{
\includegraphics[width=1.0\textwidth]{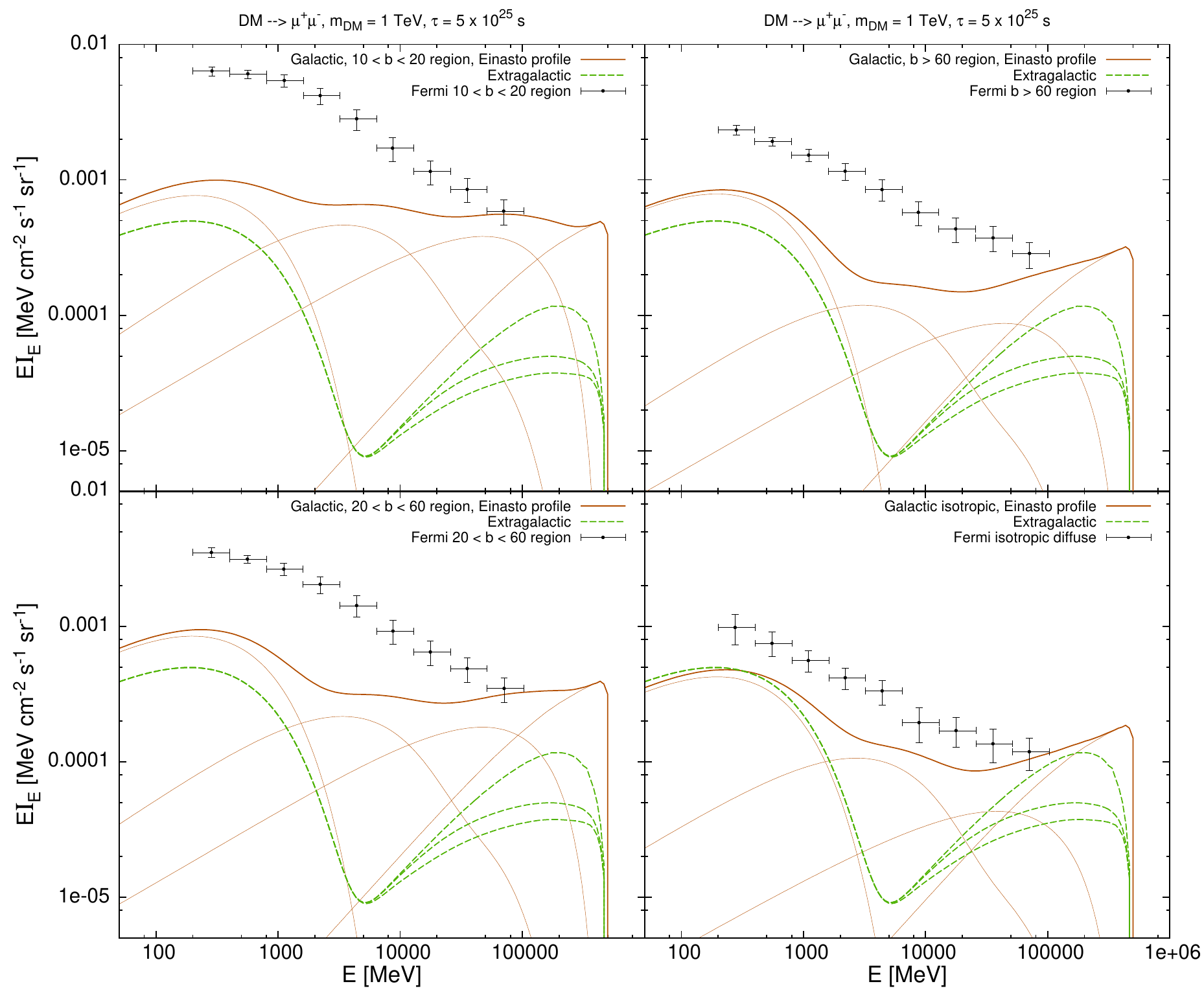}
\caption{The same as Fig.~\ref{fig1} for the decaying DM with half-life $\tau=5\times10^{25}$ s. Galactic and extragalactic spectra are given with solid and dashed lines, respectively.}
\label{fig2}}

In the following sections, where we compare the models to the Fermi-LAT diffuse gamma-ray data, we need to calculate intensities averaged over the annuli defined by ranges of the Galactic $b$ coordinate:
\begin{equation}
I_E^{b_{\min}<b<b_{\max}}=\frac{1}{2\pi(\mu_{\max}-\mu_{\min})}\cdot\int\limits^{\mu_{\max}}_{\mu_{\min}}{\rm d}\mu\int\limits_{0}^{2\pi}I_E(\mu,l){\rm d}l\,,
\end{equation}
where $\mu=\sin(b)$. In what follows we will also use an isotropic component of the Galactic DM gamma-ray emission. To extract it we find for each photon energy the direction of the minimum intensity and use the resulting minimum value as our estimate for the amplitude of the isotropic spectrum. As is probably expected, it turns out that the isotropic intensity obtained this way practically coincides with an intensity towards $b=0$, $l=\pi$, i.e. towards the direction directly away from the Galactic center. Even though the spectral shape depends on the direction one looks at, the drop in overall amplitude is almost completely driven by the rapid falloff of the DM density profile, and so irrespective of the energy, the minimum intensity is always practically towards the Galactic anticenter.

Some example Galactic spectra are shown in Figs.~\ref{fig1} and \ref{fig2} with solid red lines. In each case we have shown the total gamma-ray spectrum along with its decomposition into different components. Here the components in the order of increasing energy are: (i) IC from CMB, (ii) IC from infrared radiation, (iii) IC from stellar light, (iv) prompt gamma-ray emission. In Fig.~\ref{fig1} we have assumed $2\mu$ annihilation channel with a particle mass $m_{DM}=1$ TeV and annihilation cross-section $\cs$ $60$ times the standard thermal production value of $3\times10^{-26}$ cm$^3$s$^{-1}$. The upper and lower left-hand panels show the spectra averaged over the Galactic $10<b<20$ and $20<b<60$ regions, respectively. The right-hand panels correspond to the $b>60$ region (top panel) and to the isotropic case (bottom panel). In all of the panels the points with errorbars show the measurements of the Fermi-LAT \cite{collaboration:2010nz}. Here all the calculated Galactic models have assumed Einasto density profile. Note how the shape of the spectrum changes, particularly notice how the slight valley between the IC CMB part and the prompt part of the spectrum gets filled up, as one moves from the isotropic case towards the cases that involve more central regions, i.e. $20<b<60$ and $10<b<20$. The analogous results for the decaying DM with $\tau=5\times10^{25}$ s half-life are given in Fig.~\ref{fig2}. In comparison to the annihilating DM case, where the spectra reach $E=m_{DM}$, the spectra for the decaying DM extend to $\frac{1}{2}m_{DM}$, as expected.    

\FIGURE{
\includegraphics[width=0.75\textwidth]{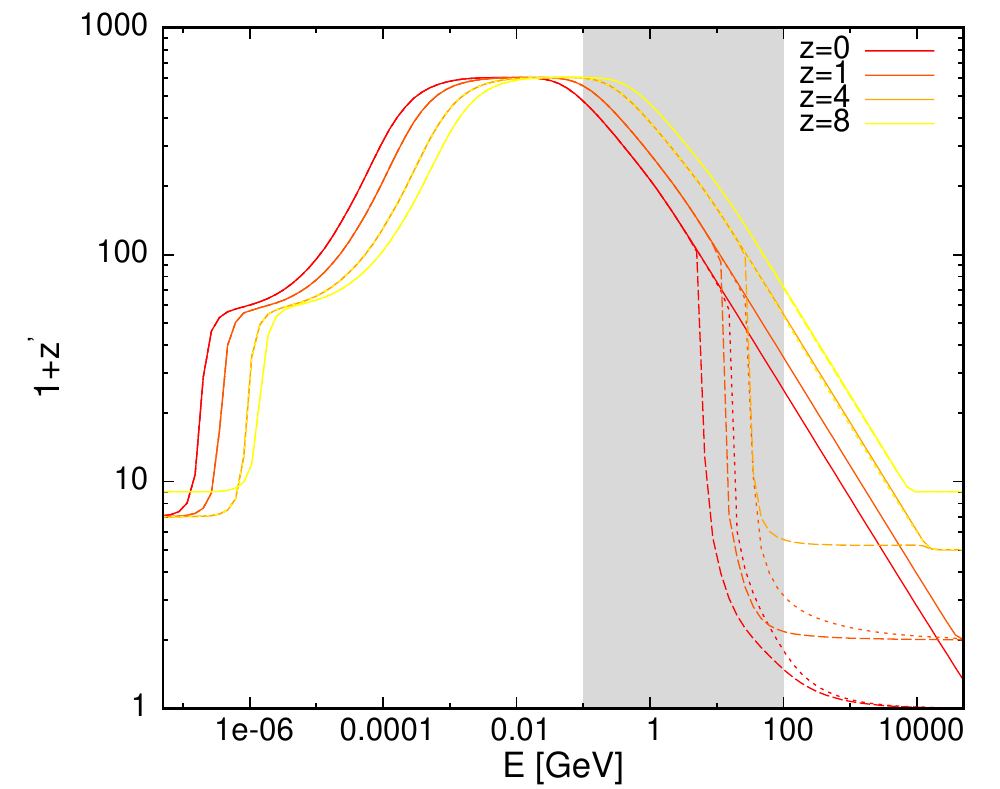}
\caption{The redshift $z^{\prime}$ where the optical depth for photons reaches unity (i.e., $\tau(E,z,z^{\prime})=1$ in Eq.~(\ref{eq_3.4}) ) for several ``observer's redshifts'': $z=0,1,4,8$. Here the energy plotted is the photon energy at redshift $z$. The solid, long-, and short-dashed curves correspond to ``no UV'', ``optimistic UV'', and ``realistic UV'' background cases, respectively. The shaded gray region gives the energy range where the Fermi-LAT measurements are done.}
\label{fig3}}

\subsection{Extragalactic contribution}
Due to the expansion of the space, even in the absence of the absorption, the intensity is no longer conserved along the line of sight. In order to calculate the extragalactic contribution to the gamma-ray intensity due to annihilating/decaying DM we have to include this cosmological dimming. Also, no longer can we neglect the absorption of gamma-rays once cosmologically large distances are involved. Thus, as in H\"utsi et al. \cite{Huetsi:2009ex}, we can write for the intensity at redshift $z$
\begin{equation}\label{eq_3.4}
I_E(z,E)=\frac{c}{4\pi}\int\limits_z^{\infty}{\rm d}z^{\prime}\frac{1}{H(z^{\prime})(1+z^{\prime})}\left(\frac{1+z}{1+z^{\prime}}\right)^3j_E(z^{\prime},E^{\prime}) \exp\left[ -\tau(E,z,z^{\prime})\right]\,.
\end{equation}
Here the factor $(H(z)(1+z))^{-1}$, where $H(z)$ is the Hubble function, converts the redshift interval to proper distance interval, the factor $((1+z)/(1+z^{\prime}))^3$ accounts for the cosmological dimming of intensities, and the function $\tau(E,z,z^{\prime})$ is the optical depth describing the absorption between redshifts $z$ and $z^{\prime}$. $E^{\prime}\equiv \frac{1+z^{\prime}}{1+z}E$ is the energy of a photon at redshift $z^{\prime}$, assuming it has energy $E$ at redshift $z$. To calculate the Hubble function we assume a flat $\Lambda$CDM ``concordance'' cosmology with $\Omega_m=0.27$ and $h=0.7$. In Fig.~\ref{fig3} we have plotted redshifts $z^{\prime}$ where $\tau(E,z,z^{\prime})=1$ for a wide range of photon energies and for several ``observer's redshifts'': $z=0,1,4,8$. Here the energy plotted is the photon energy at redshift $z$, which is the reason why the curves move to right once larger observer's redshift is taken. In the lowest energy section of the plot, before the first small plateau, the energy losses are dominated by photoionizations. Beyond that, up to the beginning of the large plateau, the Compton losses are dominating. At the large flat plateau region the energy loss is dominated by pair production on matter. In the final falling part of the curves the energy losses are determined by photon-photon pair production. To calculate the absorption coefficients for all of those processes we follow the treatment in Zdziarski \& Svensson \cite{Zdziarski:1989aa}. The energy range relevant for this paper, i.e. the range where the Fermi-LAT measurements are done, is given in Fig.~\ref{fig3}  by a shaded gray area. Thus, to calculate the optical depth we have to take into account the processes of pair production on matter and on ambient photon fields. Also, the photon-photon scattering gives some contribution in the region where the flat plateau of Fig.~\ref{fig3} turns over to falling curves. We provide a compact description of how we claculate $\tau(E,z,z^{\prime})$ in Appendix~\ref{appb}. In addition to the CMB photons we have included the UV background, which is produced in the lower redshift Universe once the first stars start to light up. In Fig.~\ref{fig3} we have shown $\tau(E,z,z^{\prime})=1$ curves for three distinct models: (i) no UV background (solid lines), (ii) ``optimistic'' UV background (long-dashed lines), where for $z>4$ we have used the UV background model of Inoue et al. \cite{Inoue:2009kd}, which is smoothly joined to lower $z$ estimates of Stecker et al. \cite{Stecker:2005qs}, (iii) ``realistic'' UV background (short-dashed lines), where we have taken into account that recent studies of blazars, e.g. \cite{Aliu:2008ay}, suggest significantly lower values for the UV photon densities than estimated in many of the previous investigations, and so we reduce the UV background level by an order of magnitude for $z>4$ and connect this model smoothly to the $z=0$ model of Stecker et al. \cite{Stecker:2005qs}. Compared to the recent UV background estimates of Gilmore et al. \cite{Gilmore:2009zb} our ``realistic'' UV model is still somewhat on the ``optimistic'' side. However, for our study these discrepancies are of minor importance, as in reality we have covered a very broad range of possibilities: any reasonable model would certainly be somewhere between the ``no UV'' and ``optimistic UV'' cases. Also, it turns out that for most realistic models the extragalactic signal is subdominant, thus leaving the details of the UV background model quite irrelevant.

In comparison to the IC calculations of the previous subsection this time we can significantly speed up our numerics, as a simple Thomson approximation turns out to be fully adequate\footnote{See Appendix~\ref{appa} for details.}.  Compared to the Galactic case where there are plenty of eV-range photons around, here we are fully dominated by very soft target photons of the CMB\footnote{Due to the strong cosmological dimming the low redshift gamma-ray intensity is mostly originating from a rather local neighborhood only, and thus one does not have to worry about CMB photons becoming energetic once $z$ increases.}.  There is also a small contribution due to more energetic UV background photons, but those give rise to only a negligible contribution to the total IC emissivity. Due to the fact that Thomson approximation turns out to be valid, we need to calculate the IC emissivity only for one fixed redshift, e.g. for $z=0$, and then for other redshifts use simple scaling properties. For the case of annihilating DM we can write down for the total emission coefficient
\begin{equation}\label{eq_3.5}
j^{\rm annih}_E(z,E)=B(z)(1+z)^6\left[\frac{1}{1+z}\bar{\jmath}^{{\rm IC}}_{z=0}\left(\frac{E}{1+z}\right)+\bar{\jmath}^{{\rm prompt}}_{z=0}(E)\right]\,,
\end{equation}
where $\bar{\jmath}^{{\rm IC}}_{z=0}$ and $\bar{\jmath}^{{\rm prompt}}_{z=0}$ are emission coefficients at $z=0$ assuming an average DM density $\bar{\rho}$. Due to the structure formation the typical value for $\rho^2$ is not simply $\bar{\rho}^2$ but is boosted by a large factor $B(z)$, i.e. $\langle \rho^2 \rangle=B(z)\bar{\rho}^2$. To calculate the boost function $B(z)$ we use the Halo Model \cite{Cooray:2002dia} of the large-scale structure, which assumes that the global DM density field can be modeled as a superposition of DM halos with universal density profiles, and whose number density is given by the mass function $n(M,z)$. The details of how we calculate $B(z)$ along with several examples can be found in our previous work \cite{Huetsi:2009ex}\footnote{Note that there is a typo in Eq.~(2) of \cite{Huetsi:2009ex}: under the integral sign there should be an additional factor of $M$.}.  Some of the results are given at the beginning of Appendix~\ref{appc}. The function $B(z)$ is highly sensitive to the choice of the concentration-mass relation $C(M)$ for the DM halos. Since the possible choices for the $C(M)$ relation already introduce a large uncertainty in the boost function $B(z)$ we have chosen to use only one form for the DM density profiles -- NFW. $B(z)$ is also strongly dependent on the lowest halo mass $M_{\min}$. As in our previous study \cite{Huetsi:2009ex} in this work we also use two $M_{\min}$ values: $10^{-9}$ and $10^{-6}$ $M_{\odot}$, which are quite typical values for the WIMP-type DM, e.g. \cite{Bringmann:2009vf,Martinez:2009jh}\footnote{As our DM has only leptonic couplings, and $M_{\min}$ is determined by the mass inside the horizon at kinematic decoupling, one could arguably reduce $M_{\min}$ still further, i.e. for the considered scenario our values for $M_{\min}$ are surely on the conservative side.}. For the concentration-mass relation $C(M)$ we use two models: (i) Maccio et al. \cite{2008MNRAS.391.1940M} model, (ii) a simple power-law $C(M)$ model. Thus, in total we have four distinct cases for the boost function $B(z)$.

For the decaying DM, as the energy input is simply proportional to the DM density we should replace $(1+z)^6$ with $(1+z)^3$ in Eq.~(\ref{eq_3.5}) and also take $B(z)\equiv 1$, i.e.          
\begin{equation}
j^{\rm decay}_E(z,E)=(1+z)^3\left[\frac{1}{1+z}\bar{\jmath}^{{\rm IC}}_{z=0}\left(\frac{E}{1+z}\right)+\bar{\jmath}^{{\rm prompt}}_{z=0}(E)\right]\,.
\end{equation}
In Fig.~\ref{fig1} we have shown extragalactic spectra for two extreme cases: (i) $M_{\min}=10^{-6}M_{\odot}$ along with Maccio et al. $C(M)$ (dashed lines), (ii) $M_{\min}=10^{-9}M_{\odot}$ and power-law $C(M)$ (dotted lines). At higher energies the curves branch into three distinct possibilities depending on the UV model applied: the highest curve corresponds to the ``no UV'' and the lowest to the ``optimistic UV'' case. The extragalactic spectra for the decaying $2\mu$ DM model are given with dashed lines in Fig.~\ref{fig2}. It is worth noting that in comparison to the Galactic spectra extragalactic curves have a prominent minimum between the IC and prompt part of the spectra, owing to the fact that outside the Galaxy, in the typical intergalactic space, we do not have a significant number density of eV-range target photons available, which could help in filling up this minimum.

\section{Constraints from Fermi-LAT data}
In this section we present the constraints on leptonically annihilating/decaying DM models using the recent gamma-ray data from the Fermi space telescope \cite{collaboration:2010nz}. We use the Fermi-LAT data for the three distinct sky regions defined by the ranges of the Galactic $b$ coordinate: $10<b<20$, $20<b<60$, $b>60$, along with the estimated isotropic diffuse component. In our calculations we use the diffuse signals only, i.e. from the Fermi-LAT measurements in the above three sky regions we have subtracted the point source contribution as given in Abdo et al. \cite{collaboration:2010nz}. It is clear that the remaining diffuse signal contains contributions from many of the known Galactic sources, and thus DM annihilation/decay can contribute only some fraction of the total signal. In this paper we have ignored those additional complications and have calculated bounds on the models that can be considered very conservative: in calculating the constraints we have demanded that in none of the energy bins the annihilation/decay signal should exceed the 1-sigma upper values for the total diffuse signal as measured by the Fermi-LAT.

In a similar manner, the isotropic diffuse component can also have several possible ``contaminating'' Galactic contributions along with extragalactic ones like AGNs\footnote{In a recent paper the AGN contribution is estimated to be around $\sim 16\%$ \cite{Collaboration:2010gq}.}, structure formation shocks \cite{Loeb:2000na}, emission from starburst galaxies \cite{Fields:2010bw} etc. However, it is not completely unrealistic that a significant fraction of the isotropic diffuse gamma-ray signal might be due to the annihilating/decaying DM. This possibility will be explored in the next section, where we show that several DM models are indeed able to provide acceptable fits to the observed isotropic diffuse component.

Our constraints (exclusion zones) on $m_{DM}-\cs$ plane for five distinct leptonic annihilation channels: $2e$, $2\mu$, $2\tau$, $4e$, $4\mu$, and for three assumed density profiles for the Galactic DM halo: Einasto, NFW, cored isothermal, are given as shaded gray regions in Fig.~\ref{fig4}. Here the annihilation cross-section $\cs$ is given in units of the standard thermal production cross-section $\cs_{\rm std} \simeq 3\times 10^{-26}$ cm$^3$s$^{-1}$. The red ellipses present the 2- and 1-sigma favored regions of Meade et al. \cite{Meade:2009iu} inferred from the $e^\pm$ measurements of the PAMELA, Fermi, and HESS (PFH) experiments. 

\FIGURE{
\includegraphics[width=0.97\textwidth]{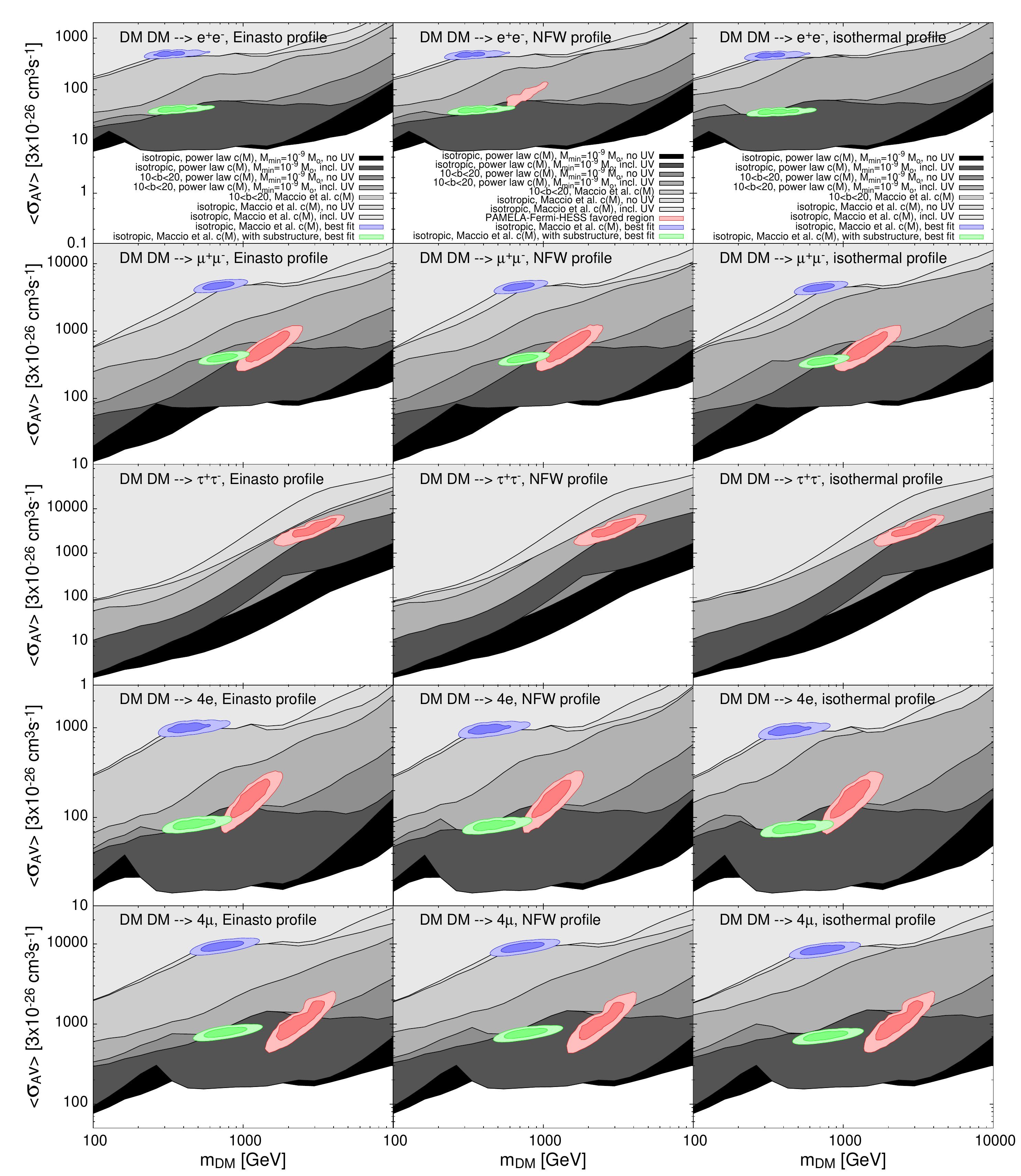}
\caption{The constraints on annihilating DM for all five channels: $2e$, $2\mu$, $2\tau$, $4e$, $4\mu$ and three DM density profiles for the Galactic halo: Einasto, NFW, cored isothermal. The gray regions represent the exclusion zones for various models as specified in the legend. The red ellipses correspond to 2- and 1-sigma favored regions of Meade et al. \cite{Meade:2009iu} inferred from the $e^\pm$ measurements of PAMELA, Fermi and HESS experiments. The blue/green ellipses represent the 2- and 1-sigma best-fit regions for the isotropic diffuse gamma-ray background for the models without/with Galactic DM halo substructure. For the case with substructure we have assumed $\rho_{{\rm sub}}^{{\rm eff}}=15\times\rho_{\odot}$.}
\label{fig4}}

In order to determine whether the constraints on Fig.~\ref{fig4} are dominated by Galactic or extragalactic component we have plotted in Fig.~\ref{fig5} the ratio of those two for one particular case: $2\mu$ channel with Einasto density profile. For all the other 14 cases of Fig.~\ref{fig4} the picture is qualitatively the same. In general, it turns out that the extragalactic signal can dominate only if power-law concentration-mass relation $C(M)$ is assumed. Once arguably more realistic Maccio et al. $C(M)$ relation is used the Galactic signal always exceeds the extragalactic one.

\FIGURE{
\includegraphics[width=0.75\textwidth]{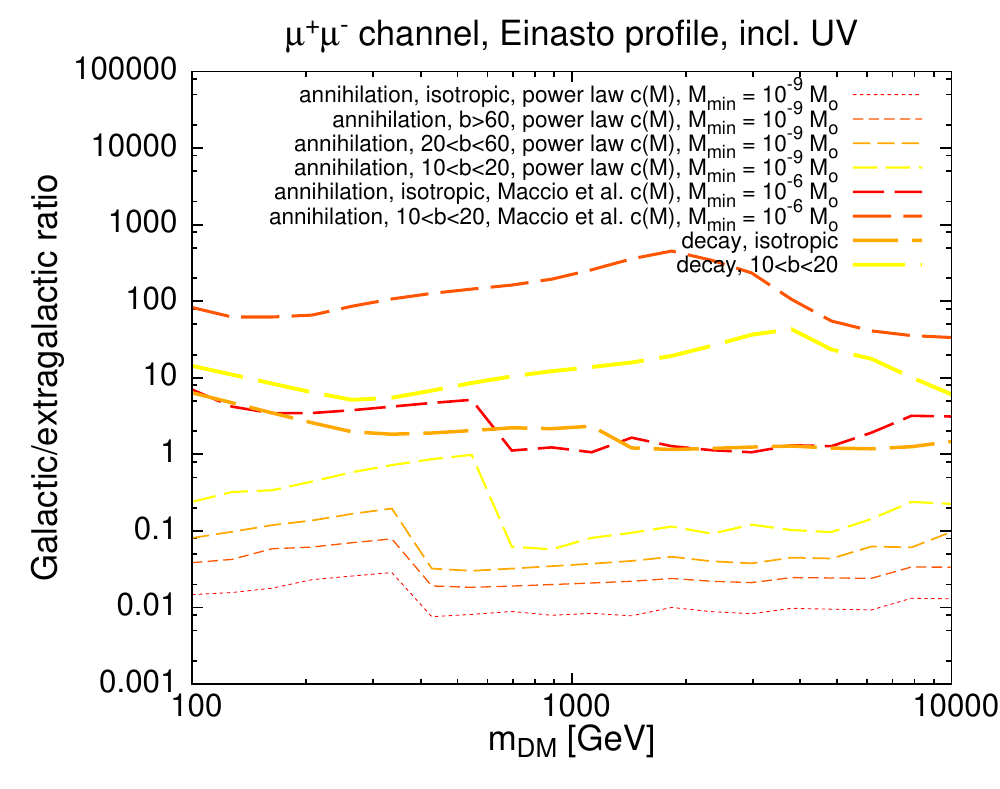}
\caption{Ratio of the Galactic to extragalactic signal for the $2\mu$ channel assuming Einasto profile for the Galactic DM halo. The extragalactic contribution dominates only if power-law $C(M)$ relation is assumed. For the other channels and density profiles the picture is qualitatively very similar.}
\label{fig5}}

In Fig.~\ref{fig4} the strongest bounds are obtained from the isotropic diffuse gamma background once the power-law $C(M)$ relation with $M_{\min}=10^{-9}M_{\odot}$ is assumed. Since in this case the extragalactic signal dominates there is a clear sensitivity to the chosen UV background model, and of course the strongest bounds correspond to the ``no UV'' case. The $10<b<20$ region gives somewhat weaker bounds, and the other regions, i.e. $20<b<60$ and $b>60$, which for clarity are not shown in the figure, give constraints between those two extremes. Once Maccio et al. $C(M)$ model is used the Galactic signal almost always dominates over the extragalactic, and thus there is evidently no sensitivity to the applied extragalactic UV model etc. Only if isotropic component is considered the Galactic and extragalactic components have a similar magnitude and so there is some sensitivity to the choice of the UV model. In contrast to the power-law case the Maccio et al. $C(M)$ gives the strongest bounds in the $10<b<20$ region and the weakest ones for the isotropic component. Again, the constraints from the $20<b<60$ and $b>60$ region fall between those extremes.

Keeping in mind that our constraints are very conservative it is clear that in the case of the power-law $C(M)$ the PFH $e^\pm$ favored regions of Meade et al. \cite{Meade:2009iu} are rather convincingly ruled out. In Fig.~\ref{fig4} we have shown the constraints for the power-law $C(M)$ with $M_{\min}=10^{-9}M_{\odot}$ only. If $M_{\min}=10^{-6}M_{\odot}$ is used instead, the bounds get weaker, but the above conclusion stays the same. On the other hand, if Maccio et al. $C(M)$ is used, we see that almost always the PFH favored regions survive except for the case of tau lepton final states which are tightly constrained due to the large fraction of the final state prompt photons.

The similar constraints for the decaying DM, for the models that are able to fit the PFH $e^\pm$ data are given in Fig.~\ref{fig6}. As can be seen from Fig.~\ref{fig5} in the case of decaying DM the Galactic signal always exceeds the extragalactic one\footnote{Note that this result disagrees with the findings of \cite{Pohl:2009qt}, where the authors claim a significantly higher contribution from the extragalactic component.}. Hence the bounds are rather insensitive to the choice of the UV model. As there are now less possibilities to consider we have shown the constraints for all three Galactic regions, along with the one from the isotropic diffuse component. It turns out that the strongest bounds are now obtained from the isotropic component and the weakest ones from the $10<b<20$ region. We see that in the $2\tau$ case the PFH $e^\pm$ favored region is mostly ruled out, while for the other two cases, i.e. $2\mu$ and $4\mu$, the exclusion zones are still below the PFH $e^\pm$ ellipses.    

\FIGURE{
\includegraphics[width=0.97\textwidth]{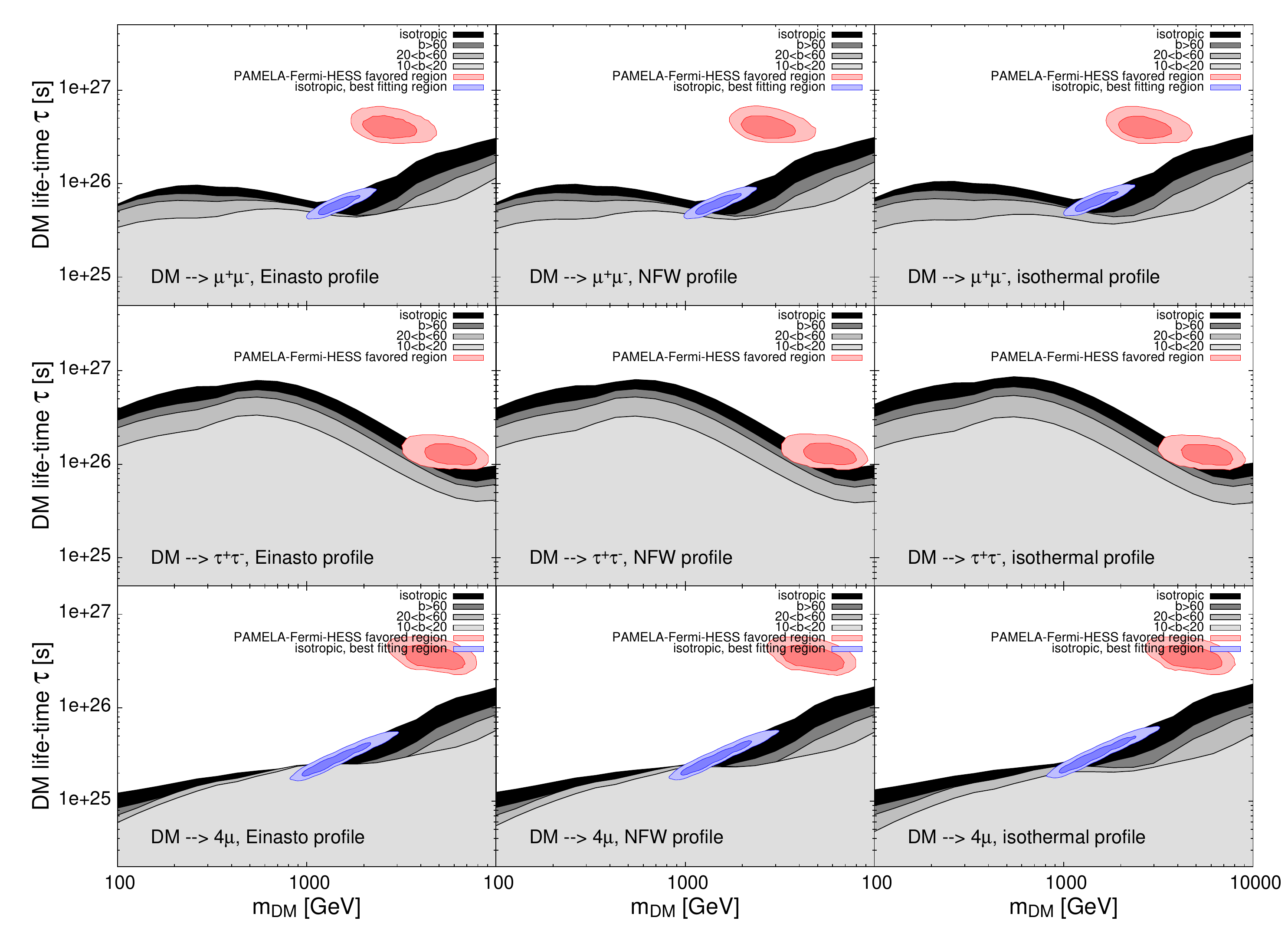}
\caption{The same as Fig.~\ref{fig4} for the decaying DM. Only the models which fit PAMELA, Fermi and HESS $e^\pm$ data are shown.}
\label{fig6}}

\section{Fitting the isotropic diffuse gamma-ray data}

\FIGURE{
\includegraphics[width=1.0\textwidth]{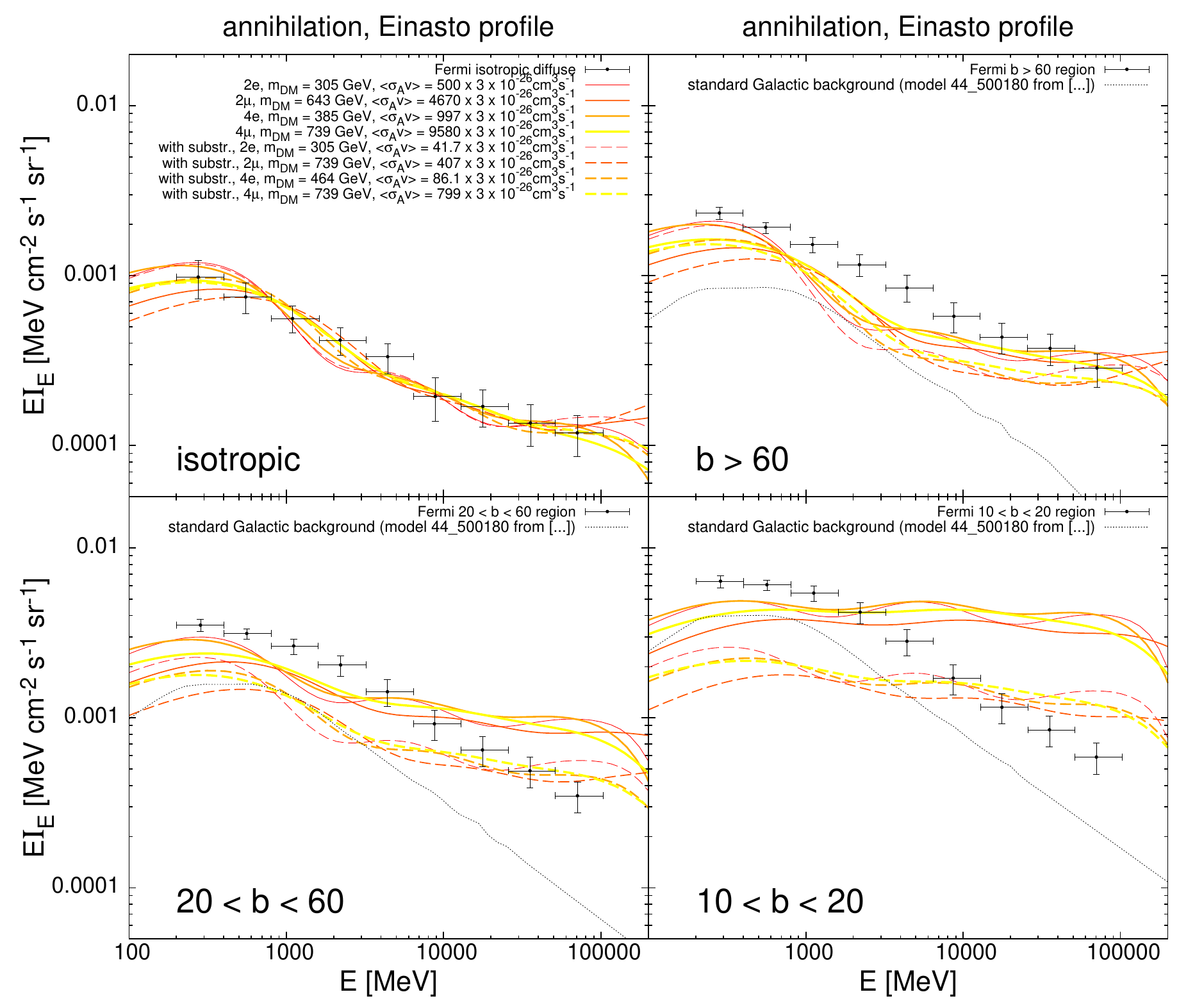}
\caption{Best fit annihilating DM models (with Einasto density profile) for the isotropic diffuse gamma-ray background (upper left panel). The solid and dashed lines represent the models without and with the Galactic substructure, respectively. For the case with substructure we have assumed $\rho_{{\rm sub}}^{{\rm eff}}=15\times\rho_{\odot}$. The averaged spectra of the same models in several Galactic regions are also shown: $b>60$ (upper right panel), $20<b<60$ (lower left panel), $10<b<20$ (lower right panel). Dotted lines show the standard Galactic diffuse background (IC + bremsstrahlung + $\pi^0$-decay) model $44\_500180$ from Strong et al. \cite{Strong:2004de}. {\bf NOTE:} The fit is performed only for the isotropic diffuse component (upper left panel). For the other regions the minimal requirement for acceptability is that models should not exceed the data points.}
\label{fig7}}

Looking at the model spectra for the isotropic component given on the lower right panels of Figs.~\ref{fig1} and \ref{fig2} one notices that the shapes are quite similar to the observed data, and so it is tempting to carry out a fitting exercise. Even though, as mentioned above, there are many possible sources like AGNs, emission from starburst galaxies, gamma-rays from structure formation shocks etc that might contribute to the isotropic diffuse gamma-ray background, it is not completely unrealistic that a large portion of this signal might be due to DM. Under that brave assumption we have carried out a fitting procedure using both annihilating and decaying DM models. 

As expected, there are indeed regions in the parameter space which provide acceptable fits. By looking at Fig.~\ref{fig1} it is clear that in order to obtain an acceptable fit for the annihilating DM one has to keep the extragalactic component as small as possible, which is achieved by using Maccio et al. $C(M)$ relation. Only the $2\tau$ channel, due to its large fraction of prompt photons, is not able to provide acceptable fits. For all the other channels the best fitting 2- and 1-sigma regions are given as blue ellipses in Figs.~\ref{fig4} and \ref{fig6}. So far we have assumed that the Galactic DM halo has a smooth density distribution characterized by density profiles $\rho(r)$ as given in Eq.~(\ref{eq_3.2}). However, in case of the CDM cosmologies we know that structure formation proceeds according to the hierarchical bottom-up scenario, and thus one would expect significant amount of substructure (subhalos) inside our Galaxy's DM halo. As discussed e.g. in Martinez et al. \cite{Martinez:2009jh}, depending on a particular substructure model, these subhalos might potentially boost the smooth halo annihilation signal by a factor of $\sim 1-20$. In our calculations we will assume a simple model for the halo substructure, where the subhalos follow the main halo's density profile and also that subhalo mass distribution function along with concentration-mass relation is independent of the location inside the main halo. Under those assumptions, as shown in Appendix~\ref{appc}, the gamma-ray emissivity due to halo substructure scales as $\propto \rho(r)$, i.e. similar to the case with decaying DM. As shown in Appendix~\ref{appc} under the above simplifying assumptions the specific details of the substructure model can be absorbed into a single location-independent parameter $\rho_{{\rm sub}}^{{\rm eff}}$, allowing the calculations to proceed practically the same way as for the case without substructure, only $\rho^2(r)$ needs to be replaced by $\rho_{{\rm sub}}^{{\rm eff}}\cdot\rho(r)$. If one takes $\rho_{{\rm sub}}^{{\rm eff}}=\rho_{\odot}=0.4$ GeV cm$^{-3}$ then the signal from substructures turns out to be approximately equal to the main halo contribution, and in loose terms one can say that the substructure boost (1 + substructure signal / main halo signal) is approximately equal to $2$. In order to lower the blue best-fit regions of Fig.~\ref{fig4} close to the PFH favored red ellipses one has to have substructure boost factors of order $\sim 10$. In Fig.~\ref{fig4} the green regions correspond to the case where $\rho_{{\rm sub}}^{{\rm eff}}=15\times\rho_{\odot}$.    

The best fitting annihilating DM model spectra for the $2e$, $2\mu$, $4e$, and $4\mu$ channels are given on the upper left-hand panel of Fig.~\ref{fig7}. The solid and dashed lines represent the models without and with the Galactic substructure, respectively. For the case with substructure we have assumed $\rho_{{\rm sub}}^{{\rm eff}}=15\times\rho_{\odot}$. In the other panels we have plotted the spectra for the same models in three Galactic regions: $b>60$ (upper right panel), $20<b<60$ (lower left panel), and $10<b<20$ (lower right panel). We see that although the models can nicely fit the isotropic diffuse component, the ones without substructure clearly exceed the observational data once more central regions of Galaxy are included, i.e. $20<b<60$ and $10<b<20$ regions. Once substructure is included those problems with central regions start to disappear. In reality, the substructure does not need to simply follow the density profile of the main halo, but in the central regions the destruction probability for subhalos might significantly increase. This helps in lowering the central signal with respect to the contribution from the outskirts of the halo, and thus making the gamma-ray emission profile shallower. Note that the profile can also be made shallower if one allows the intermediate particle $\phi$ (see Section~\ref{sec1}) to have a sufficiently long lifetime, so that it can move out of the initial place of production before decaying, which would effectively correspond to the smoothing of the gamma-ray emission profile\footnote{Of course, the acceptable range of lifetimes is limited, as in the most radical case, where the lifetime of $\phi$ gets very large, the Galactic signal becomes very diluted and the typical IC spectrum would be generated on extragalactic photon fields, instead.}.  Following this line of thought it is not hard to imagine the case where the problems with the central regions in Fig.~\ref{fig7} disappear altogether. Thus, it seems quite possible to have annihilating DM models, which: (i) are compatible with the PFH $e^\pm$ data, and (ii) provide acceptable fits to the Fermi-LAT isotropic diffuse background, without being in clear conflict with the other Fermi-LAT gamma-ray measurements. As the modern N-body simulations are only capable of resolving DM substructures that are several orders of magnitude larger than the mass scales relevant for the current problem, one is necessarily limited to this type of qualitative arguments only.

The analogous results for the decaying DM are given in Fig.~\ref{fig8}. Here in contrast to the annihilation case (without substructure) we see that the best fitting models of the upper left-hand panel do not exceed the observational data points in other regions. However, the blue best-fit regions in Fig.~\ref{fig6} are quite far from the PFH $e^\pm$ ellipses.
   
\FIGURE{
\includegraphics[width=1.0\textwidth]{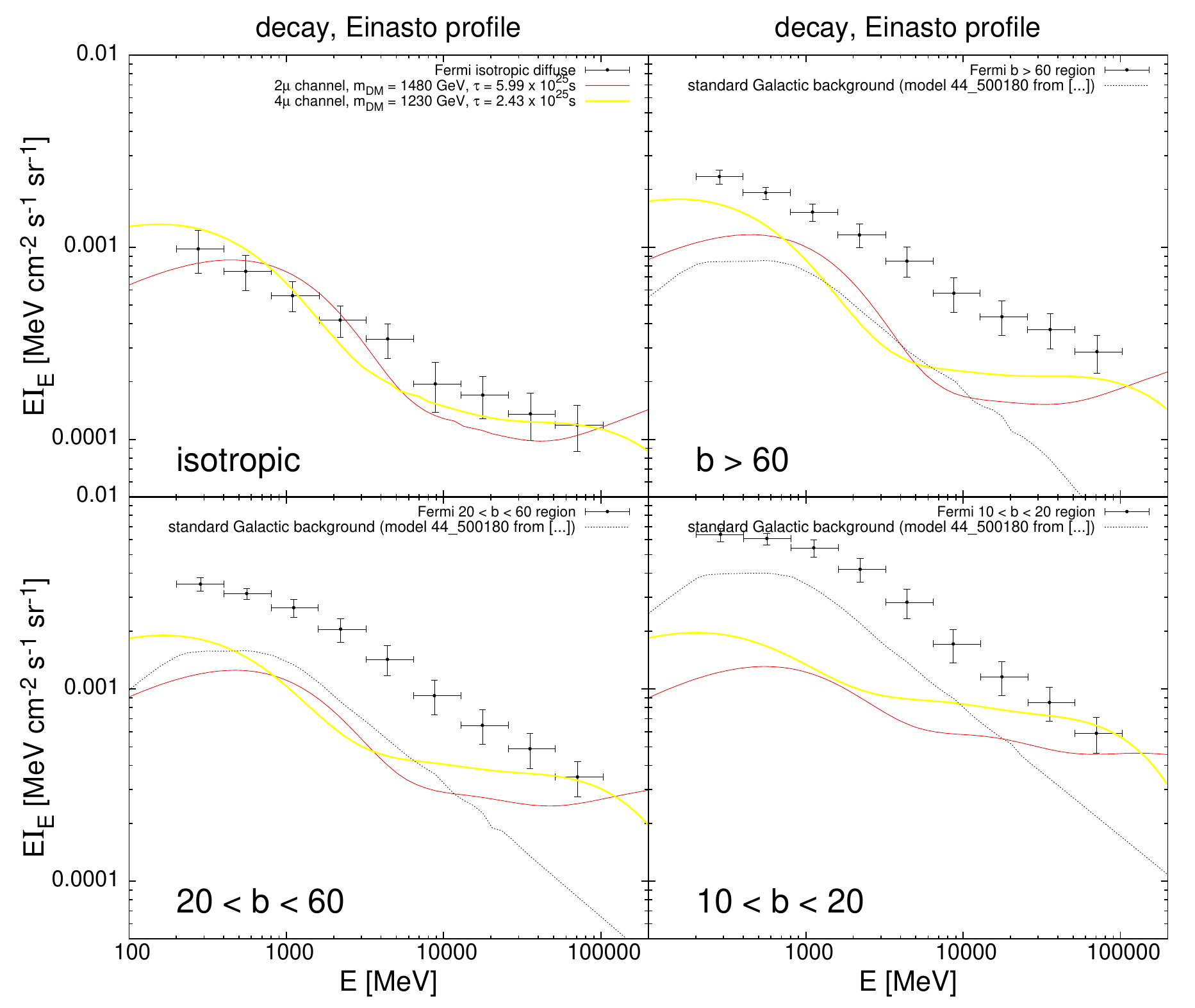}
\caption{The same as Fig.~\ref{fig7} for decaying DM.}
\label{fig8}}

\section{Inclusion of the standard Galactic background}
So far we have considered only DM-induced gamma-ray signal. The minimum requirement for the DM models to be valid is not to exceed the observational data points in Figs.~\ref{fig7} and \ref{fig8}. However, it is clear that diffuse gamma-ray background has other contributions. Here we consider the standard Galactic background model $44\_500180$ from Strong et al. \cite{Strong:2004de}. In Figs. ~\ref{fig7} and \ref{fig8} with dotted lines we have plotted this standard Galactic diffuse background model, which consists of IC, bremsstrahlung, and $\pi^0$-decay components. In Figs. ~\ref{fig9} and \ref{fig10} we have added this background to the DM-induced signal.

It is quite remarkable that the models, which provide the best fits to the isotropic diffuse data, automatically get quite close to the observed data points in other Galactic regions, once the standard background model of Strong et al. \cite{Strong:2004de} is included. It is evident that by a relatively modest tuning of the Galactic background model satisfactory fits in all of the regions should be possible. The only exception is the annihilating DM case for the region $10<b<20$, where at the highest energies the models are above the observational data points, even before the standard Galactic background is added. But this, as explained earlier, could in principle be alleviated by, e.g., increasing the dominance of the substructure at the outer halo regions, this way effectively reducing the steepness of the resulting gamma-ray profile.   

\FIGURE{
\includegraphics[width=1.0\textwidth]{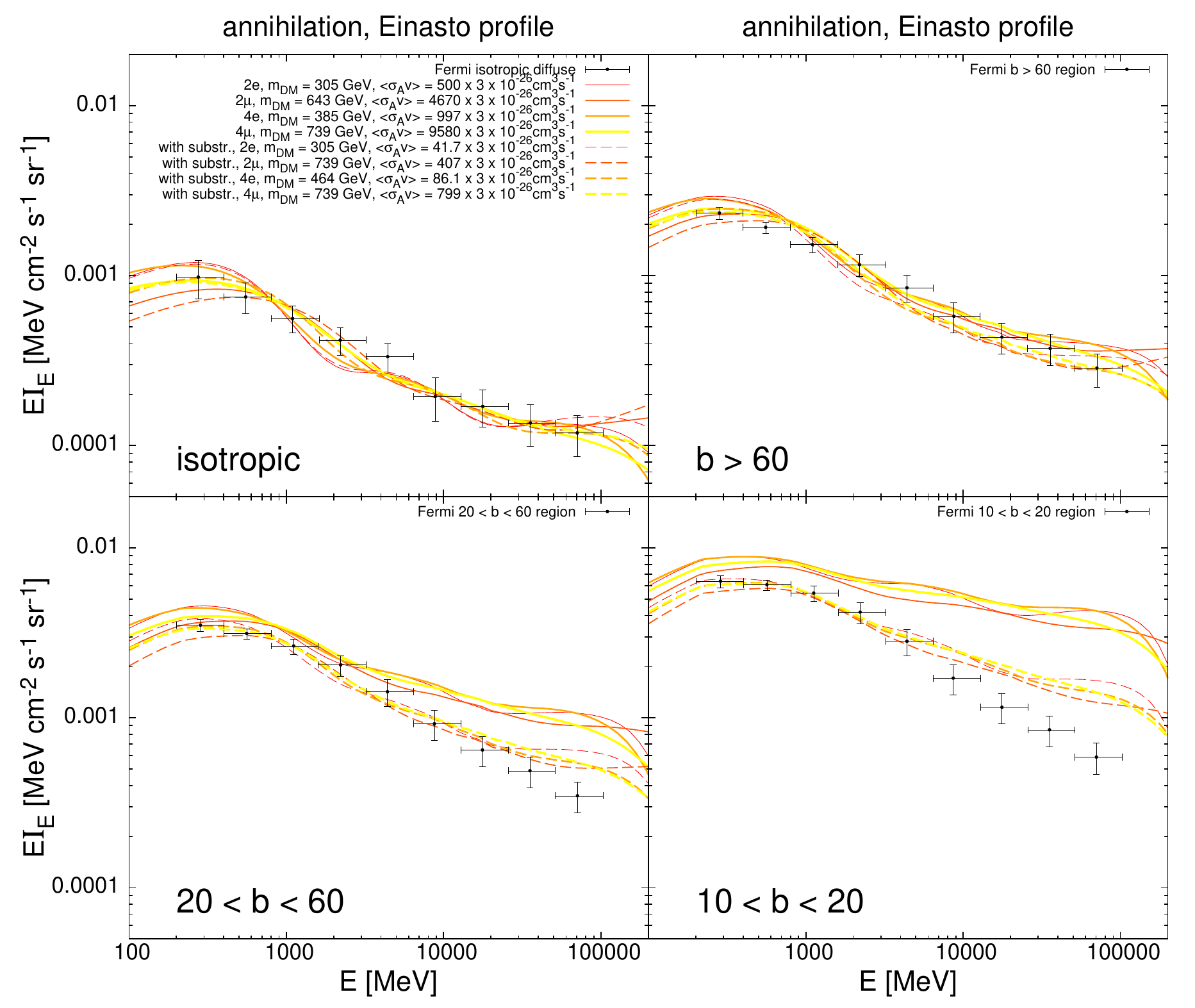}
\caption{The same as Fig.~\ref{fig7} with the standard Galactic background added to the DM signal.}
\label{fig9}}

\FIGURE{
\includegraphics[width=1.0\textwidth]{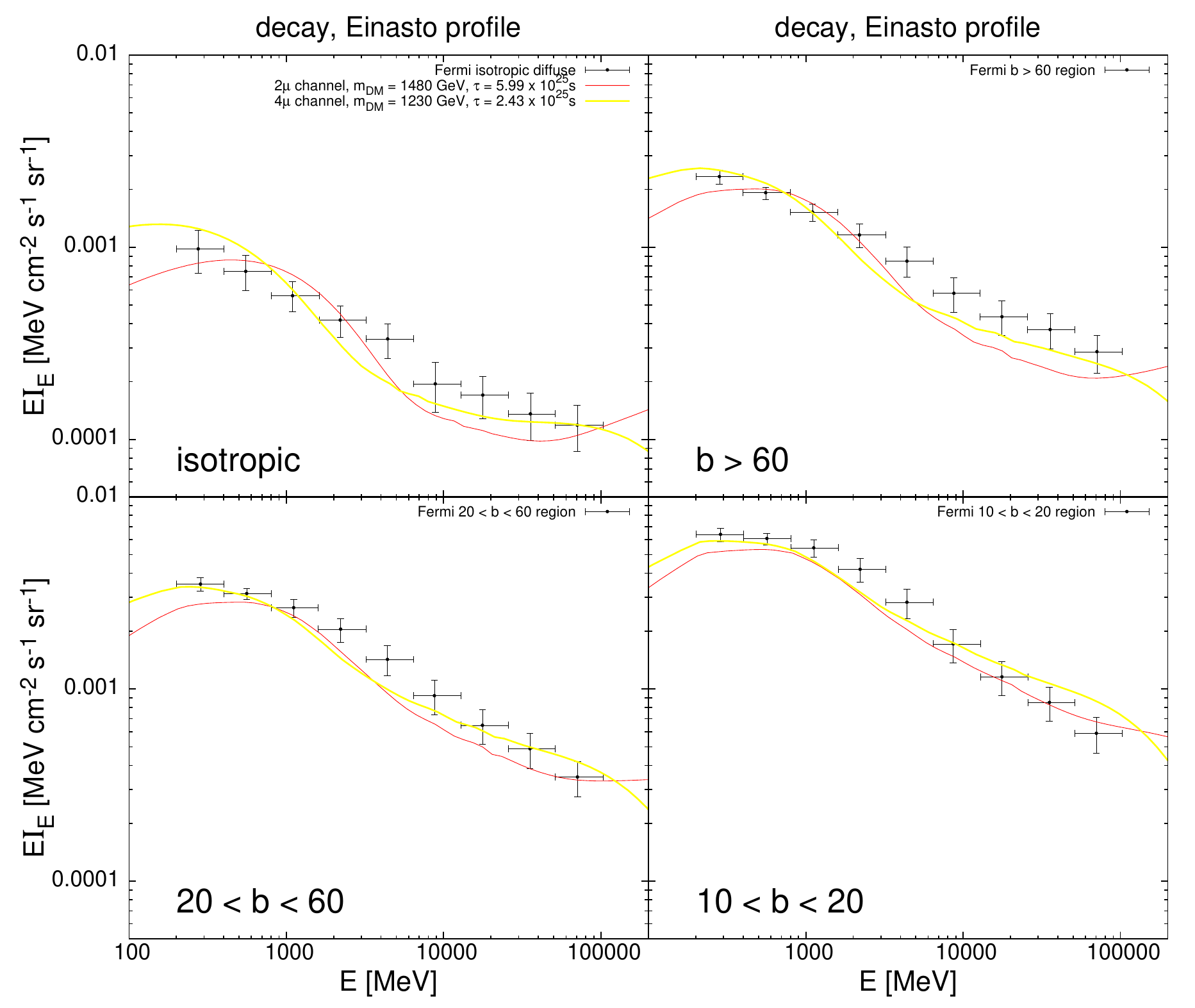}
\caption{The same as Fig.~\ref{fig8} with the standard Galactic background added to the DM signal.}
\label{fig10}}

\section{Summary and discussion}
In this paper we have used the gamma-ray data from the Fermi-LAT space telescope to place constraints on leptonically annihilating and decaying DM models. In particular, we have used the diffuse gamma-ray data for the Galactic regions $b>60$, $20<b<60$, and $10<b<20$, along with the isotropic diffuse component as estimated in Abdo et al. \cite{collaboration:2010nz}. Thus, in our calculations we do not consider the very central regions of the Galaxy, as the closest we get is $10$ degrees from the center. This new Fermi data has been also used in some of the other analogous studies. In contrast to the present study, where we calculate both prompt and IC parts of the gamma-ray spectra, the authors of \cite{Abazajian:2010sq} have focused only on the prompt emission part. In Abdo et al. \cite{Abdo:2010dk} the authors have paid a significant amount of attention to the extragalactic contribution, but their Galactic signal does not seem to be very realistic, as the IC and prompt components of the spectra are clearly separated, which hints that their model for the ISRF used in the IC calculations, has probably neglected the stellar radiation and the infrared radiation from the Galactic dust. In comparison to our paper, were we have investigated both decaying and annihilating DM, the above studies consider only the annihilating case. The preliminary Fermi data has been used to constrain DM models in Meade et al. \cite{Meade:2009iu}, in Cirelli et al. \cite{Cirelli:2009dv}, and in Papucci \& Strumia \cite{Papucci:2009gd}. Compared to our previous study \cite{Huetsi:2009ex} where we focused only on extragalactic signals, this time we have included the contribution from our own Galactic DM halo. The most important difference between our analyses and the previous papers is that we have also performed fits of the diffuse gamma-ray data and compared the results with the 
preferred DM parameters explaining the PFH $e^\pm$ anomalies.

Here are our main conclusions:
\begin{itemize}

\item For the annihilating DM models the extragalactic signal can dominate only if power-law concentration-mass relation $C(M)$ is assumed. Once arguably more realistic Maccio et al. $C(M)$ relation is used the Galactic signal always exceeds the extragalactic one. Due to those reasons our calculated bounds on annihilating DM have significant dependence on the assumed extragalactic UV background model (which strongly effects the level of prompt emission) only in the case of the power-law $C(M)$ relation. In the case of decaying DM, where the signal scales proportionally to the DM density, and thus is not very sensitive to the details of the structure formation model, the Galactic signal is always stronger than the extragalactic one. As a result the calculated bounds in this case are rather insensitive to the choice of the extragalactic UV model.

\item The derived constraints are very conservative, as the only thing we have required is that none of the models should exceed the measured diffuse signals in any of the observed energy bins. It is clear that in reality those diffuse signals can have various Galactic and extragalactic contributions other than the potential contribution from the annihilating/decaying DM. Our constraints for the annihilating and decaying DM models are given in Figs.~\ref{fig4} and \ref{fig6}, respectively. The red elliptical regions in those figures show the PFH $e^\pm$ favored regions of Meade et al. \cite{Meade:2009iu}. It turns out that for the power-law $C(M)$ relation the PFH favored regions are quite convincingly ruled out. On the other hand, if Maccio et al. $C(M)$ is used, we find that almost always the PFH favored regions easily survive except for the tau final states that produce too many prompt gamma-rays.  For the decaying DM only the $2\tau$ channel is significantly constrained, while for the other two cases, i.e. $2\mu$ and $4\mu$, the exclusion zones are still below the PFH $e^\pm$ ellipses.

\item Once realistic models for the ISRF are used with all the relevant components: (i) stellar light, (ii) infrared radiation from dust, (iii) CMB, it turns out that the resulting gamma-ray spectra from the Galactic annihilating/decaying DM halo can be remarkably close to a seemingly featureless power law over the relevant energy range, as seen  in Figs.~\ref{fig1} and \ref{fig2}. Thus, there might be a possibility that those models provide an acceptable fit to the isotropic diffuse gamma-ray data if extragalactic contribution can be kept low, which is surely true for the Maccio et al. $C(M)$ relation. Indeed, this turns out to be the case: in Figs.~\ref{fig4} and \ref{fig6} the blue ellipses show the 2- and 1-sigma best-fit regions for the scenario neglecting Galactic DM halo substructure. 
Only the $\tau$ lepton final states fail to give reasonable fits to the data due to hard prompt photons. However,  there is no overlap between the gamma-ray and PFH best-fit regions. 
Several best fitting spectra corresponding to this case are plotted with solid lines on the upper left-hand panels of Figs.~\ref{fig7} and \ref{fig8} for the annihilating and decaying DM, respectively. It is clear from Fig.~\ref{fig7} that the best fitting annihilating models are in conflict with other measurements from other regions, especially with the ones from the $10<b<20$ Galactic region. However, in the case of the decaying DM there is no immediate conflict (i.e. the model spectra do not exceed the data points) with the other measurements, as seen from Fig.~\ref{fig8}. 

\item All the above results did not include the possible contribution from the Galactic DM substructure. We have devised a simple model for the halo substructure (see Appendix~\ref{appc}), where all the specific model details can be absorbed into a single effective substructure density parameter $\rho_{{\rm sub}}^{{\rm eff}}$. The inclusion of substructure has a twofold effect: (i) it leads to the increase of the total amplitude of gamma-ray emissivity, (ii) it makes the spatial emission profiles shallower compared to the smooth halo case. Due to those reasons the inclusion of substructure lowers the blue best-fit regions of Fig.~\ref{fig4}, and at the same time helps to reduce the conflict with the measurements of other Galactic regions seen on the lower panels of Fig.~\ref{fig7}. The green best-fit regions of Fig.~\ref{fig4} and the dashed lines of Fig.~\ref{fig7} correspond to the case with $\rho_{{\rm sub}}^{{\rm eff}}=\rho_{\odot}=0.4$ GeV cm$^{-3}$, which in loose terms represents the case with substructure boost factor $\simeq 15$. Those levels of substructure boosts are indeed possible, as discussed, e.g., in Martinez et al. \cite{Martinez:2009jh}. If we give up our simplifying assumption that substructure follows the density profile of the main halo, and allow an increased subhalo destruction rate in the center, it should be possible to make the spatial emission profiles even shallower, and thus potentially remove the remaining small conflict on the lower right-hand panel of Fig.~\ref{fig7}. The emission profiles can also be made shallower if one allows the intermediate particle $\phi$ (see Section~\ref{sec1}) to have a sufficiently long lifetime, so that it can move significantly away from the initial place of birth before decaying, which would effectively reduce the steepness of the resulting gamma-ray profile.

It is clear that diffuse gamma-ray background has other contributions. In this work we consider the standard Galactic background model $44\_500180$ from Strong et al. \cite{Strong:2004de}. In Figs. ~\ref{fig9} and \ref{fig10} we have added this background to the DM-induced signal. It is quite remarkable that the models, which provide the best fits to the isotropic diffuse data, automatically get quite close to the observed data points in other Galactic regions, once the standard background model of Strong et al. \cite{Strong:2004de} is included. It is evident that by a relatively modest tuning of the Galactic background model satisfactory fits in all of the regions should be possible.

We conclude that it seems quite realistic to have annihilating DM models, which simultaneously: (i) are compatible with the PFH $e^\pm$ data, and (ii) provide acceptable fits to the Fermi-LAT isotropic diffuse background, without being in clear conflict with the other Fermi-LAT gamma-ray measurements.
The preferred DM annihilation final states consistent with this analysis and the PFH fits are $2\mu,$ $4\mu$ and $4e$. 

\item In our Galactic IC calculations we have neglected the effect of electron diffusion, i.e. in our treatment the IC spectrum is produced ``on spot''. For more precise treatment one should solve the diffusion-loss equation for electrons to determine the electron energy distribution, which is used as an input in IC calculations. Some estimates of Papucci \& Strumia \cite{Papucci:2009gd} have shown that by including this diffusion effect the IC signal gets reduced by a factor of around $2-3$ depending on the Galactic region. However, the full calculation of electron diffusion is quite complicated with rather large uncertainties due to, e.g., the knowledge of the magnetic fields. Even in our simple treatment there is a large uncertainty due to the assumed ISRF model. We have used the latest ISRF model from the GALPROP code, which is based on results from Porter \& Strong \cite{Porter:2005qx}. However, in their study the authors do not provide any errors for the estimated ISRF. It is clear that their results are highly model dependent, as from the direct measurements we can only get the ISRF in our location, while the values at other Galactic locations can only be obtained by solving the radiative transfer problem, which clearly involves several model-dependent assumptions. According to Strong et al. \cite{Strong:2004de} a factor of two uncertainty for the ISRF is quite possible. Due to those reasons we feel that the accurate treatment for the $e^\pm$ diffusion problem can probably wait before there are better models with realistic uncertainties available for the ISRF. Unfortunately, except for rare examples of \cite{Porter:2005qx} and \cite{Strong:1998fr}, the problem of determining the ISRF as a function of Galactic position has not gained much attention so far. 
%In principle, the estimate for the position-dependent ISRF could be readily obtained as a side product from the studies like \cite{2006A&A...459..113M} aiming to determine Galactic gas, dust and stellar distributions from observational data.  

\item In Introduction we referred to the other constraints on annihilating
or decaying DM. Compared to the CMB constraints (see e.g. \cite{Slatyer:2009yq,Huetsi:2009ex,Cirelli:2009bb}) the bounds derived in this paper can be stronger or weaker, depending on what type of concentration-mass relation $C(M)$ one assumes. If Maccio et al. \cite{2008MNRAS.391.1940M} $C(M)$ is used the bounds are generally weaker, while the power-law $C(M)$ leads to stronger constraints than obtainable from the CMB measurements. The CMB constraints, being mostly determined by the free electron fraction at redshifts $100 \lesssim z \lesssim 1000$ \cite{Huetsi:2009ex}, are quite insensitive to the uncertainties of the nonlinear structure formation and other complicated low-redshift astrophysical phenomena. In that respect they can be considered significantly more robust than the diffuse gamma-ray bounds derived in this paper. 

The constraints from the Galactic center radio and gamma-ray measurements are generally stronger than ours, however those are hampered by the obscurity of the DM density profile, ISRF and magnetic field in the very central region (see, e.g. \cite{Bergstrom:1997fj,Dodelson:2007gd,Bertone:2008xr,Bringmann:2009ca,Crocker:2010gy}). For example, there is no clear understanding on how much various physical processes involving baryonic component can impact on the the central cusp of the DM profile (see e.g. \cite{deBlok:2009sp} for a recent review). 

The constraints derived from the Galactic dwarf spheroidal galaxies and the nearby galaxy clusters are comparable or slightly weaker than our restrictions \cite{Essig:2009jx,Abdo:2010ex,FermiDM:2010aa}.

\end{itemize}

\subsection*{Note added}
When the research presented in this work was completed, Ref.~\cite{Lin:2010fb} appeared that also performs fits of the cosmic ray and gamma-ray data. Although their approach and aim of the paper is different form ours, the main conclusions that only DM annihilations can potentially provide a satisfactory fit agree with each other.

\acknowledgments
We thank  Alessandro Strumia, Marco Cirelli and Mario Kadastik for discussions and our Referee for useful comments and suggestions. This work was supported by the ESF Grants 8499, 8090, 8005, 7146, by SF0690030s09 and by EU  FP7-INFRA-2007-1.2.3 contract No 223807. GH acknowledges the support from the EstSpacE fellowship.

\appendix
\section{Emission coefficients for DM annihilation and decay}\label{appa}
The total gamma-ray emissivity $j_E$ $[$eV~cm$^{-3}$s$^{-1}$eV$^{-1}]$ for the leptonically annihilating/decaying DM models considered in this paper can be given as a sum of IC and prompt parts
\begin{equation}
j_E(E) = j_E^{{\rm IC}}(E) + j_E^{{\rm prompt}}(E)\,,
\end{equation}
where the prompt emission part is trivial to calculate once the input photon spectra from PYTHIA simulations are available. In this appendix we focus on the IC calculations. In what follows we assume that the IC spectrum is generated ``on spot'', allowing us to use one fixed value for the DM density, $\rho$, as well as for the number density of target photons with energy $E \ldots E+{\rm d}E$, $n_E(E)$. The IC emissivity can then be expressed as
\begin{equation}      
j_E^{{\rm IC}}(E) = \left\{
\begin{array}{cl}
\frac{\cs}{2m_{DM}}\rho^2\ &{\rm for\ annihilating\ DM}\\
\rho/\tau\ &{\rm for\ decaying\ DM}
\end{array}
\right\}\times\int\limits_{m_e}^{m_{DM}}{\rm d}E_e\frac{K_1(E,E_e)}{K_2(E_e)}\int\limits_{E_e}^{m_{DM}}{\rm d}\widetilde{E}_e\frac{{\rm d}N_e}{{\rm d}\widetilde{E}_e}\,,
\end{equation}
where $\frac{{\rm d}N_e}{{\rm d}E_e}$, which is calculated with PYTHIA, is the spectrum of produced $e^\pm$ per annihilation/decay. The functions $K_1(E,E_e)$ and $K_2(E_e)$ are given as:
\begin{itemize}
\item Full Klein-Nishina case
\begin{eqnarray}
K_1(E,E_e)&=&E\int\limits_{\frac{1}{4\gamma^2}}^{1}{\rm d}x\left[1 - \frac{1}{4\gamma^2x(1-\kappa)}\right]\frac{n_E(\widetilde{E}(x))}{x}\nonumber \\
&\times& \left[2x\ln x+ \left(1+2x+\frac{1}{2}\frac{\kappa^2}{1-\kappa}\right)(1-x)\right]\,,\\
K_2(E_e)&=&\int\limits_{0}^{\infty}{\rm d}E\,K_1(E,E_e)\,,
\end{eqnarray}
where
\begin{equation}
\gamma\equiv\frac{E_e}{m_e}\,\,,\quad\kappa\equiv\frac{E}{E_e}\,\,,\quad\widetilde{E}(x)\equiv\frac{1}{4\gamma^2x}\frac{E}{1-\kappa}\,.
\end{equation}
\item Thomson approximation ($\kappa\ll 1$)
\begin{eqnarray}
K_1(E,E_e)&=&E\int\limits_{0}^{1}{\rm d}x\frac{n_E(\widetilde{E}(x))}{x}\left[2x\ln x+(1+2x)(1-x)\right]\,,\\
\widetilde{E}(x)&\equiv&\frac{E}{4\gamma^2x}\,,\\
K_2(E_e)&=&\int\limits_{0}^{\infty}{\rm d}E\,K_1(E,E_e)=\left(\frac{4}{3}\gamma^2\right)^2\int\limits_{0}^{\infty}{\rm d}E\,n_E(E)E\,.
\end{eqnarray}
\end{itemize}

\section{Absorption coefficients and optical depth for gamma photons}\label{appb}
As described in the main text, the processes relevant for the absorption of gamma-rays in the energy range probed by the Fermi-LAT are: (i) pair production on matter, (ii) photon-photon scattering, and (iii) pair production on ambient photon fields. At those energies by far the dominating one amongst the three is the pair production on photon fields. Here we provide a brief summary of results needed to calculate the optical depth $\tau(E,z,z^{\prime})$ in Eq.~(\ref{eq_3.4}) for gamma-ray photons between redshifts $z$ and $z^{\prime}$. For for further details see \cite{Zdziarski:1989aa} and \cite{1988ApJ...335..786Z}. In what follows we express the gamma photon energy in units of electron rest mass, i.e. $\epsilon\equiv \frac{E}{m_ec^2}$.
\begin{itemize}
\item The absorption coefficient at redshift $z$ for pair production on neutral matter with mass fractions of hydrogen $X=0.75$ and helium $Y=0.25$ can be approximated as
\begin{eqnarray}
\alpha_{{\rm mat-pair}}(\epsilon,z)&\simeq&\alpha_0(1+z)^3\ln\left(\frac{513\epsilon}{\epsilon+825}\right),\quad\epsilon>6\,,\\
\alpha_0\simeq 5.3n_e^0\alpha_fr_0^2&\simeq& 2.05\times 10^{-9}\left(\frac{\Omega_b}{0.045}\right)\left(\frac{h}{0.7}\right)^2\,{\rm Mpc}^{-1}\,,
\end{eqnarray}
where the average electron number density at $z=0$ is
\begin{equation}
n_e^0 \simeq 2.17\times 10^{-7} \left(\frac{\Omega_b}{0.045}\right)\left(\frac{h}{0.7}\right)^2\,{\rm cm}^{-3}\,.
\end{equation}
Here $\Omega_b$ is the density parameter for baryons, $h$ is the Hubble parameter, $\alpha_f= 7.29735\times10^{-3}$ is the fine structure constant, and $r_0= 2.8179\times10^{-13}$ cm the classical electron radius.

However, below redshift $z\simeq6$ the Universe is known to be reionized \cite{Fan:2006dp}. The absorption coefficient for pair production on fully ionized matter can be approximately given as
\begin{eqnarray}
\alpha_{{\rm ionmat-pair}}(\epsilon,z)&\simeq&\alpha_0(1+z)^3\left[\ln(2\epsilon)-\frac{109}{42}\right],\quad\epsilon \gg 1\,,\\
\alpha_0\simeq \frac{20}{3}n_e^0\alpha_fr_0^2&\simeq& 2.58\times 10^{-9}\left(\frac{\Omega_b}{0.045}\right)\left(\frac{h}{0.7}\right)^2\,{\rm Mpc}^{-1}\,.
\end{eqnarray}
\item The absorption coefficient for photon-photon scattering at redshift $z$ can be given by
\begin{eqnarray}
\alpha_{{\rm \gamma\gamma-scat}}(\epsilon,z)&=&\alpha_0(1+z)^6\epsilon^3\,,\\
\alpha_0= \frac{4448\pi^4}{455625}\frac{\alpha_f^4\Theta_0^6}{\lambda_c}&\simeq& 3.23\times 10^{-31}\left(\frac{T_0}{2.725\,{\rm K}}\right)^6\,{\rm Mpc}^{-1}\,,
\end{eqnarray}
where $\Theta_0\equiv \frac{k_BT_0}{m_ec^2}$ is the CMB temperature at $z=0$ in electron rest mass units, and $\lambda_c= 2.4263\times 10^{-10}$ cm is the electron Compton wavelength.
\item For the photon-photon pair production, in addition to the soft CMB photons, we have to include photons of the extragalactic UV background, which are produced in the low redshift Universe once the first stars start to light up\footnote{The influence of those UV photons on the previously discussed photon-photon scattering process, which itself is a weak effect, is completely negligible.}.  As there are no simple analytical forms available for the UV background spectrum, we have to use full numerical treatment for the photon-photon pair production process. The absorption coefficient at redshift $z$ for photon-photon pair production can be given as
\begin{eqnarray}
\alpha_{{\rm \gamma\gamma-pair}}(\epsilon,z)&=&\frac{\lambda_c^2\alpha_f^2}{4\pi}\int\limits_{1/\epsilon}^{\infty}{\rm d}\tilde{\epsilon}\,n(\tilde{\epsilon},z)\frac{\phi(v)}{(\epsilon\tilde{\epsilon})^2}\,,\\
\phi(v)=\frac{1+2v+2v^2}{1+v}\ln w &-& \frac{2\sqrt{v}(1+2v)}{\sqrt{1+v}} -\ln^2 w + 2\ln^2 (1+w) + \nonumber \\
&+&4\,{\rm Li}_2\left(\frac{1}{1+w}\right) - \frac{\pi^2}{3}\,,\\
v\equiv\epsilon\tilde{\epsilon}-1\ge0\,,\quad w&\equiv&\frac{\sqrt{1+v} + \sqrt{v}}{\sqrt{1+v} - \sqrt{v}}\,,
\end{eqnarray}
where $n(\epsilon,z)=n_{{\rm CMB}}(\epsilon,z)+n_{{\rm UV}}(\epsilon,z)$ is the number density of target photons, and ${\rm Li}_2$ is the dilogarithm function.
\item Now we can write for the total absorption coefficient
\begin{equation}
\alpha(\epsilon,z)=\left\{
\begin{array}{rl}
\alpha_{{\rm mat-pair}}(\epsilon,z)\ &{\rm if}\ 6\lesssim z \lesssim 1000\\
\alpha_{{\rm ionmat-pair}}(\epsilon,z)\ &{\rm if}\ z \lesssim 6
\end{array}
\right\} + \alpha_{{\rm \gamma\gamma-scat}}(\epsilon,z) + \alpha_{{\rm \gamma\gamma-pair}}(\epsilon,z)\,,
\end{equation}
and for the optical depth $\tau(E,z,z^{\prime})$ between redshifts $z$ and $z^{\prime}$
\begin{equation}
\tau(E,z,z^{\prime})=c\int\limits_{z}^{z^{\prime}}{\rm d}\tilde{z}\frac{\alpha(\tilde{\epsilon},\tilde{z})}{H(\tilde{z})(1+\tilde{z})}\,,
\end{equation}
where $\tilde{\epsilon}=\frac{1+\tilde{z}}{1+z}\frac{E}{m_ec^2}$. To calculate the Hubble function $H(z)$ we assume a flat $\Lambda$CDM model with $\Omega_m=0.27$ and $h=0.7$.
\end{itemize}

\section{Halo substructure: a simple model}\label{appc}
In our previous paper \cite{Huetsi:2009ex} we showed that within the Halo Model of the large-scale structure the typical value of $\rho^2$ at redshift $z$ gets boosted by a factor $B(z)$ above the average DM density squared, i.e. $\langle \rho^2(z)\rangle=B(z)\bar{\rho}^2(z)$, where 
\begin{equation}
B(z)=1+\frac{\Delta_c}{\bar{\rho}}\int\limits_{M_{\min}}^{\infty}{\rm d}M M\frac{{\rm d}n}{{\rm d}M}(M,z)f_C\left[C(M,z)\right]\,.
\end{equation}
Here $\Delta_c$ is the overdensity (with respect to the mean DM density) used for defining halo masses in simulations, with a typical value of $\Delta_c\simeq 200$. $M_{\min}$ is the lower cutoff for the halo mass, $\frac{{\rm d}n}{{\rm d}M}$ is the halo mass function, and the function $f_C$, which depends on the halo concentration-mass relation, is given for the NFW halos as
\begin{equation}
f_C(C)=\frac{C^3}{9}\left[1-\frac{1}{(1+C)^3}\right]\left[\ln(1+C)-\frac{C}{1+C}\right]^{-2}\,.
\end{equation}

This calculation looked at the contribution from the ensemble of halos, but did not include the substructure inside each of these. It is clear that similar calculation can be repeated for one particular halo plus an ensemble of subhalos inside it. One just has to replace the halo mass function with subhalo mass function $\frac{{\rm d}n}{{\rm d}M}_{{\rm sub}}$. Thus, for the substructure boost factor as a function of distance from the Galactic center we can write (we do not include +1 as this corresponds to the contribution from the mean smooth DM component)
\begin{equation}
B_{{\rm sub}}(r)=\frac{\Delta_c}{\bar{\rho}}\int\limits_{M_{{\rm sub},\min}}^{\infty}{\rm d}M_{{\rm sub}} M_{{\rm sub}}\frac{{\rm d}n}{{\rm d}M}_{{\rm sub}}(M_{{\rm sub}},r)f_C\left[C(M_{{\rm sub}},r)\right]\,.
\end{equation}
Now we make assumptions that the subhalo mass distribution function $f_{{\rm sub}}(M_{{\rm sub}})$ \\($\int {{\rm d}M}_{{\rm sub}}f_{{\rm sub}}(M_{{\rm sub}})\equiv1$) does not depend on Galactocentric radius $r$, and that the spatial distribution of subclumps follows the density profile $\rho(r)$ of the main halo. We also assume that the concentration-mass relation $C(M_{{\rm sub}},r)$ does not depend on $r$. Under those simplifying assumptions we can express the subhalo mass function via
\begin{equation}
\frac{{\rm d}n}{{\rm d}M}_{{\rm sub}}(M_{{\rm sub}},r)=\frac{f_Mf_{{\rm sub}}(M_{{\rm sub}})}{M_{{\rm sub}}}\rho(r)\,.
\end{equation}
Here $f_M$ is the ratio of the total substructure mass and the main halo mass. For the substructure boost factor we can then write
\begin{equation}
B_{{\rm sub}}(r)=f_M \Delta_c \frac{\rho(r)}{\bar{\rho}}\int\limits_{M_{{\rm sub},\min}}^{\infty}{\rm d}M_{{\rm sub}} f_{{\rm sub}}(M_{{\rm sub}})f_C\left[C(M_{{\rm sub}})\right]\,.
\end{equation}
Here the only $r$ dependent term is $\rho(r)$.

In our calculations we need $\langle \rho_{{\rm sub}}^2(r)\rangle$, which can be be given by
\begin{equation}
\langle \rho_{{\rm sub}}^2(r)\rangle = B_{{\rm sub}}(r)\bar{\rho}^2\equiv \rho_{{\rm sub}}^{{\rm eff}}\cdot\rho(r)\,,
\end{equation}
where the effective substructure density
\begin{equation}
\rho_{{\rm sub}}^{{\rm eff}} \equiv f_M \Delta_c \bar{\rho}^2\int\limits_{M_{{\rm sub},\min}}^{\infty}{\rm d}M_{{\rm sub}} f_{{\rm sub}}(M_{{\rm sub}})f_C\left[C(M_{{\rm sub}})\right]\,.
\end{equation}
Thus, under the above simplifying assumptions all the model details can be absorbed into one $r$ independent parameter $\rho_{{\rm sub}}^{{\rm eff}}$. If we take $\rho_{{\rm sub}}^{{\rm eff}}=\rho_{\odot}=0.4$ GeV cm$^{-3}$ then the gamma-ray signal from the main halo and from substructures turn out to be of very similar magnitude. However, note that the spatial profiles of the signals are quite different: the gamma-ray emissivity profile from substructures is similar to the case of the decaying DM, as in both cases $j_E(r)\propto \rho(r)$.

We conclude that  the inclusion of DM substructure will generally: 
(i) increase the total amplitude of gamma-ray emissivity, (ii) make the spatial emission profiles shallower compared to the smooth halo case. 

\bibliographystyle{JHEP}
\bibliography{refs}

\end{document}